\documentclass[
    aip,
    jcp,
    reprint,
    a4paper
]{revtex4-2}

\usepackage{graphicx}
\usepackage[version=4]{mhchem}
\usepackage{siunitx}
\usepackage{booktabs}
\usepackage{tabularx}
\usepackage[hidelinks]{hyperref}

\DeclareSIUnit{\angstrom}{\textup{\smash{Å}\vphantom{A}}}

\begin{document}

\title{Interpretable machine learned predictions of adsorption energies at the metal--oxide interface}

\author{Marius Juul Nielsen}
\affiliation{Center for Interstellar Catalysis, Department of Physics and Astronomy, Aarhus University, Aarhus C, Denmark}

\author{Luuk H. E. Kempen}
\affiliation{Center for Interstellar Catalysis, Department of Physics and Astronomy, Aarhus University, Aarhus C, Denmark}

\author{Julie de Neergaard Ravn}
\affiliation{Center for Interstellar Catalysis, Department of Physics and Astronomy, Aarhus University, Aarhus C, Denmark}

\author{Raffaele Cheula}
\affiliation{Center for Interstellar Catalysis, Department of Physics and Astronomy, Aarhus University, Aarhus C, Denmark}

\author{Mie Andersen}
\email{mie@phys.au.dk}
\affiliation{Center for Interstellar Catalysis, Department of Physics and Astronomy, Aarhus University, Aarhus C, Denmark}

\date{\today}

\begin{abstract}
    The conversion of \ce{CO2} to value-added compounds is an important part of the effort to store and reuse atmospheric \ce{CO2} emissions. Here we focus on \ce{CO2} hydrogenation over so-called inverse catalysts: transition metal oxide clusters supported on metal surfaces. The conventional approach for computational screening of such candidate catalyst materials involves a reliance on density functional theory (DFT) to obtain accurate adsorption energies at a significant computational cost. Here we present a machine learning (ML)-accelerated workflow for obtaining adsorption energies at the metal--oxide interface. We enumerate possible binding sites at the clusters and use DFT to sample a subset of these with diverse local adsorbate environments. The data set is used to explore interpretable and black-box ML models with the aim to reveal the electronic and structural factors controlling adsorption at metal--oxide interfaces. Furthermore, the explored ML models can be used for low-cost prediction of adsorption energies on structures outside of the original training data set. The workflow presented here, along with the insights into trends in adsorption energies at metal--oxide interfaces, will be useful for identifying active sites, predicting parameters required for microkinetic modeling of reactions on complex catalyst materials, and accelerating data-driven catalyst design.
\end{abstract}

\maketitle

\section{Introduction}

The conversion of \ce{CO2} into industrially relevant value-added compounds has received attention as part of a wider initiative to lower atmospheric \ce{CO2} concentrations~\cite{kapiamba2022inverse, wang2023co2, ye2019co2}. The source of \ce{CO2} in this process could be sequestered atmospheric \ce{CO2} obtained via carbon capture technologies, allowing for carbon-neutral manufacturing of C1 compounds such as methanol~\cite{olah2011anthropogenic}. 

The current conventional method of methanol production involves the use of synthesis gas derived from fossil fuels, a mixture of \ce{CO} and \ce{H2} with a small concentration (\qty{3}{\percent}) of \ce{CO2}, in the presence of a \ce{Cu}/\ce{ZnO}/\ce{Al2O3} catalyst~\cite{guil2019methanol}. The synthesis of methanol in this way involves balancing three coupled equilibrium reactions~\cite{schittkowski2018methanol}.
\begin{align}
    \ce{CO2 + 3 H2 &<--> CH3OH + H2O} \label{rea:co2hydrog}
    \\
    \ce{CO2 + H2 &<--> CO + H2O} \label{rea:rwgs}
    \\
    \ce{CO + 2 H2 &<--> CH3OH} \label{rea:cohydrog}
\end{align}

Isotopic labeling has shown that the predominant carbon source in the methanol formed is \ce{CO2}, indicating that the hydrogenation of \ce{CO2} is the main reaction for methanol synthesis~\cite{chinchen1987isotope}. Hydrogenation of \ce{CO2} (reaction~\eqref{rea:co2hydrog}) is aided by the high concentration of \ce{CO} in the syngas reactant mixture, which suppresses the competing reverse water--gas shift (RWGS) reaction (reaction~\eqref{rea:rwgs}). This in turn lowers the fraction of water formed at equilibrium, which could otherwise inhibit methanol production by shifting the equilibrium of reaction~\eqref{rea:co2hydrog} and cause sintering of the catalyst~\cite{wu2001stability}. 

When utilizing a \ce{CO2}-rich feed the influence of the RWGS reaction is significant, leading to lower methanol yields. The issue is made worse at higher temperatures where the the endothermic RWGS reaction is thermodynamically favorable~\cite{guil2019methanol}. This necessitates the development of novel catalysts that are stable, exhibit high activity and methanol selectivity and can operate at lower temperatures where the RWGS reaction is less favorable~\cite{guil2019methanol}. To this end, inverse catalyst configurations, characterized by metal oxide clusters on a metal support, are a promising class of catalyst materials for \ce{CO2} hydrogenation~\cite{kapiamba2022inverse}. Studies of \ce{CO2} hydrogenation on a range of inverse catalysts have shown that they can be more active than their conventional metal on oxide counterparts~\cite{wu2020inverse, senanayake2016hydrogenation, graciani2014ceria}.

In the search for new catalytic materials, hereunder inverse catalysts, computational screening approaches to narrow the pool of candidate materials have become an important aid to experimental studies~\cite{broadbelt2000applications}. Microkinetic models are often employed in these screening studies to evaluate reaction kinetics and estimate the catalytic activity of candidate materials. Key quantities in the construction of these models are the adsorption energies of intermediates and activation energies of reaction steps at specific sites on the catalyst surface. Often, activation energies can be estimated from adsorption energies via Brønsted--Evans--Polanyi (BEP) scaling relations~\cite{bronsted1928bep,evans1936bep,michaelides2003bep}. 

Reliable values for adsorption energies are typically obtained from DFT calculations. On the surface of single-component crystalline materials such as metals or oxides, the high symmetry means that there are typically only at most a handful of unique binding sites, limiting the amount of required DFT evaluations. However, in the case of inverse catalysts, the oxide nanoclusters come in many different sizes and compositions and typically present structures with a lack of symmetry, resulting in a large space of unique binding configurations on the surface. The use of DFT in exploring this space introduces a prohibitive computational cost in the screening process.

Machine learning (ML) tools are increasingly incorporated in materials screening in order to alleviate this computational bottleneck by forgoing most of the DFT calculations. Computational savings are made by employing ML models for the prediction of adsorption energies of reaction intermediates at active sites on the surface~\cite{deimel2020active, andersen2019beyond, ulissi2017address, li2017high}. When targeting energy predictions, ML models are typically trained in a supervised learning approach on a subset of DFT-evaluated energies and thereafter used to predict adsorption energies on similar unseen binding configurations.

Another application of ML tools is to discover underlying trends and correlations in large data sets and thereby gain physical insights into the factors controlling the target for the predictions; in this case the adsorption energy. ML models such as neural networks and Gaussian process regression are often so complex that determining the trends learned by the model, and which inform their predictions, is difficult. There are however models which, by virtue of their simplicity or built-in metrics, offer interpretability~\cite{andersen2019beyond,esterhuizen2022interpretable,xin2023}. These interpretable models offer a vehicle for gleaning insight into the underlying structure--property correlations that inform adsorption energy prediction. Insight which may lead to the development of new knowledge that can accelerate the discovery of promising candidate catalytic materials. 

In this work, we employ four different ML models with varying degrees of interpretability for the prediction of adsorption energies. We focus on the adsorption of formate (\ce{HCOO}), since previous studies on various different catalysts have found it to be a key intermediate with high stability in the \ce{CO2} hydrogenation reaction mechanism~\cite{kattel2017,frei2018,cheula2024}.
Specifically, we consider the binding of formate on active sites of an ensemble of inverse catalyst materials: the combinations of \ce{In_yO_x} and \ce{Zn_yO_x} nanoclusters on \ce{Au(111)}, \ce{Cu(111)}, and \ce{Pd(111)} surfaces. The best ML model obtained on the combined data set from all materials can predict binding energies for unseen formate binding configurations with a root-mean-square-error (RMSE) of about \qty{0.17}{\eV}, which is accurate enough to be useful for screening purposes. Feature importance analysis highlights the importance of the work function of the catalyst surface, a quantity that is also experimentally assessable, and which is highly correlated with the oxygen content of the cluster. Further important features are related to the electronic properties of the binding site atoms, e.g., the Pauling electronegativity and features related to the projected density of states.

\section{Methods}

\subsection{Adsorption Energy Calculations}\label{sec:meth-ads_e}

The adsorption energy, $E_{\mathrm{ads}}$, is calculated with DFT as the energy of the inverse catalyst system with formate adsorbed, minus the energy of the clean inverse catalyst and the energy of the gas-phase formate.
\begin{equation}
    E_{\mathrm{ads}} = E_{\mathrm{sys}} - (E_{\mathrm{clean}} + E_{\ce{HCOO}})
    \label{eq:ads_e}
\end{equation}

Details on the settings used in the DFT calculations are given in Section~\ref{sec:meth-dft}. Clean inverse catalyst structures are taken from the results of our previous global structure optimization search~\cite{kempen2025inverse}, where six different inverse catalysts are studied: the combinations of \ce{In_yO_x} and \ce{Zn_yO_x} clusters on \ce{Au(111)}, \ce{Cu(111)}, and \ce{Pd(111)} surfaces. For each system there are structures for 30 \ce{M_yO_x} stoichiometries (M being Zn or In) with $3\leq y\leq 7$ and $0\leq x\leq y$, giving 180 stoichiometries in total.
In this present work, we consider only the stoichiometries that were stable at some oxygen chemical potential within the range considered in our previous \textit{ab initio} thermodynamics (AITD) free energy analysis \cite{kempen2025inverse}. For each of the stable stoichiometries all structures within \qty{0.1}{\eV} of the global minimum energy structure are included in this work.

\subsection{Binding Site Enumeration and Sampling}\label{sec:meth-bindingsites}

\begin{figure*}
    \centering
    \includegraphics{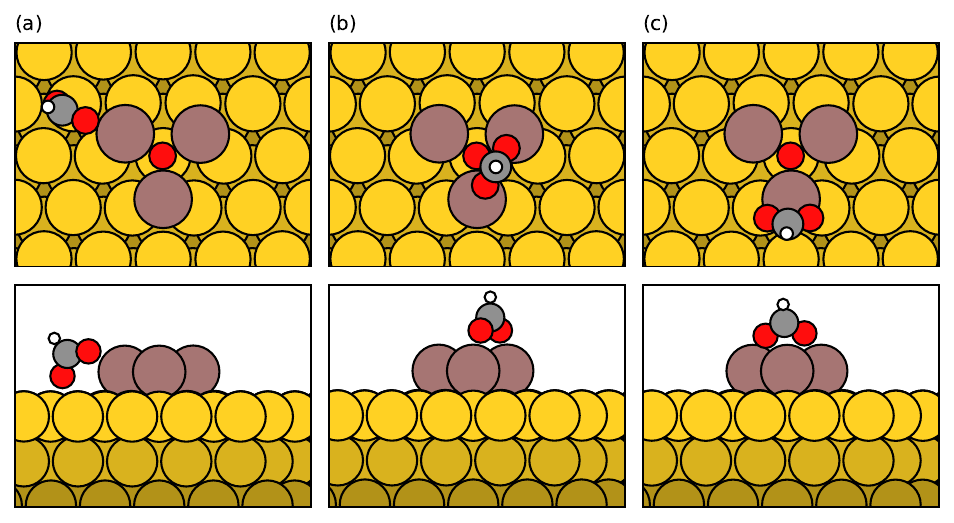}
    \caption{ Visualizations of the types of binding sites considered in the site enumeration. (a) Bridge, where formate bridges between a cluster and surface metal atom at the interface. (b) Bridge, where formate bridges over two metal atoms of the cluster. (c) On-top, where both \ce{O} atoms of formate bind to the same cluster metal atom.}
    \label{fig:bindingsite_ex}
\end{figure*}

Structures of formate bound to clusters are obtained by relaxation of a subset of binding configurations generated by enumeration over binding sites. The binding sites on clusters are identified by recognition of on-top and bridge sites on the metal oxide clusters. Examples of these are shown in Figure~\ref{fig:bindingsite_ex}. Bridge sites are characterized by the oxygen atoms being bound to two separate atoms, at least one of which must be a cluster metal atom. On-top sites are characterized by the oxygen atoms of the formate binding to the same cluster metal atom.  At each identified binding site a formate is placed, yielding a set of initial binding configurations.

From the set of all possible initial binding configurations, we sample a subset for DFT evaluation. This is done by fingerprinting each initial binding configuration with a Smooth Overlap of Atomic Positions (SOAP) descriptor~\cite{bartok2013,himanen2020dscribe,caro2019optimizing} (parameters are given in Section~S1.2 of the supplementary material). Formate atoms are chosen as centers for the SOAP descriptor to capture the local environment of the formate. Farthest point sampling (FPS) of initial binding configurations in SOAP space is then done to select the subset---using the configuration closest to the mean SOAP vector as a seed---ensuring a diverse representation of local environments.
In the FPS procedure, the point in SOAP space with the largest minimum distance to the set of previously sampled points is iteratively selected. In this work, we make use of a pre-computed distance matrix calculated with the \texttt{pairwise\_distances} method from scikit-learn~\cite{sklearn}.

DFT relaxation of the sampled initial binding configurations yields a set of relaxed structures and corresponding adsorption energies.

\subsection{Density Functional Theory Calculations}\label{sec:meth-dft}

The DFT calculations were carried out using the GPAW package~\cite{mortensen2005, enkovaara2010, mortensen2024}, which implements the projector augmented-wave method~\cite{blochl1994} in conjunction with the atomic simulation environment (ASE) package~\cite{larsen2017}. Wave functions were represented using a plane-wave (PW) basis, with a cutoff energy of \qty{400}{\eV} for all systems. 
For the real-space representation of the wave functions, a grid spacing of \qty{0.2}{\angstrom} was used.
The exchange--correlation functional was approximated using the Perdew--Burke--Ernzerhof (PBE) functional~\cite{perdew1996} with D4 dispersion corrections~\cite{caldeweyher2017,caldeweyher2019,caldeweyher2020}. Metallic supports are represented by a 4-layered (\numproduct{7x7}) fcc(111) slab model with \qty{18}{\angstrom} of vacuum between slabs. A dipole correction was included~\cite{bengtsson1999dipole}. Orbital occupancies were set using a Fermi--Dirac distribution with a width of \qty{0.1}{\eV}, except for \ce{Pd}-supported systems where a width of \qty{0.25}{\eV} is used.

Structure relaxations were performed using the Broyden--Fletcher--Goldfarb--Shanno (BFGS) algorithm with an atomic force convergence criterion of \qty{0.05}{\electronvolt\per\angstrom}. Atoms of the formate, cluster, and top surface layer were allowed to relax, while the remaining slab atoms were fixed. The Brillouin zone was sampled using only the $\Gamma$ point during relaxation. Once relaxed, structures were refined with a subsequent single-point calculation using a (\numproduct{2x2}) Monkhorst--Pack $k$-point grid~\cite{monkhorst1976}.

The clean inverse catalyst structures were evaluated with the same DFT settings as their adsorption structure counterparts. The gas-phase formate molecule was evaluated at the $\Gamma$ point with the same PW cutoff energy used in the cluster calculations and with spin-polarized DFT to account for unpaired valence electrons.

\subsection{Binding Site Features}\label{sec:meth-feats}

The predictive performance of adsorption energy models is underpinned by the features included in the data they are trained on. In this work we are interested in developing computationally efficient models which can predict adsorption energies at all sites on the cluster and cluster--surface interface from a single DFT calculation of the clean structure. Therefore we only include features obtained from the clean structures. Global features, describing the entire cluster, as well as local features of the specific binding sites are included. Features can further be categorized as either stoichiometric, geometric, or electronic. Stoichiometric features are included to describe the atomic species present at the cluster or binding site, such as the ratio of metal to oxygen atoms in the cluster or the number of nearest neighbor oxygen atoms to the binding site. Geometric features are incorporated to describe the local arrangement of atoms near the binding site, such as the mean distance to nearest-neighbor atoms. Angular-resolved local order parameters for angular quantum number values of \numrange{1}{6} are also included to capture information on the cluster geometry at the binding site~\cite{steinhardt1983}.

Electronic features are included to describe the electronic structure of the cluster or binding site in the clean structure. The considered global electronic features are the density of states (DOS) at the Fermi level and the work function ($\phi$). At the binding site atoms the Bader charge is calculated using the algorithm by \textcite{henkelman2006bader}. Several features are included to describe properties of the projected DOS (pDOS) at the binding site, which can capture qualitative information on the chemical bonding of binding site atoms to the cluster/surface. Among these pDOS features are the sp- and d-band projected centers and widths, as well as the parameters of Gaussian curves fitted to the pDOS. In the case of the graph-based model, we also include the SOAP descriptor of the clean structure atoms in the set of features. Further details on the features used in the training of adsorption energy models can be found in Section~S1 of the supplementary material. 

\subsection{Adsorption Energy Models}

In this work we focus on ML models that can be trained on small data sets. The key trait distinguishing the selected model types is the degree to which their predictions can be interpreted to provide physical insight into the structure--property correlations underlying adsorption energies.

The first three models employ a representation consisting of a vector of features describing the active site. In the case of sites with two binding site atoms, we calculate the values for site-specific features by taking the mean of the feature value for the two atoms. The last model employs a graph representation of the catalyst surface.

\subsubsection{RBF-GPR}

A Gaussian process regression (GPR) model was used as a black-box model, offering no physical interpretation of its predictions. The scikit-learn~\cite{sklearn} GPR implementation was employed with a radial basis function (RBF) kernel. Hyperparameters were tuned using a limited memory BFGS optimizer with bound constraints on hyperparameter values. Target binding energy values were $z$-normalized to obtain higher predictive accuracy.

\subsubsection{XGBoost}

Extreme Gradient Boosting (XGBoost) is a machine learning algorithm that employs gradient boosting to train data-efficient models~\cite{xgboost}. XGBoost iteratively builds an ensemble of decision trees, where each subsequent decision tree is trained to correct the error of the ensemble. XGBoost models have two key hyperparameters which determine their complexity; the number of decision trees to train and the maximum allowed depth of the decision trees. 

Interpretability of XGBoost models is facilitated by feature importance scores, which quantify the contribution of each feature to the predictive accuracy of the model. The importance score of a feature is measured through the gain, which quantifies the improvement in accuracy brought about when the feature is used in a split.

\subsubsection{SISSO}\label{sec:sisso}

The most interpretable model type employed is the Sure Independence Screening plus Sparsifying Operator (SISSO), which seeks to construct an analytical expression for a descriptor to predict the quantity of interest~\cite{sisso}. Descriptors learned by SISSO take the form shown in Equation~\eqref{eq:sisso_ex}, where the predicted adsorption energy, $E_{\mathrm{ads}}$, is given by a linear expansion with free coefficients $C_i$ and features $f_i$. There are two hyperparameters which determine the complexity of SISSO models; the dimension $M$, which determines the number of terms in the expansion, and the rung $\Phi_N$ which determines the complexity of features. 
\begin{equation}
    E_{\mathrm{ads}} = \sum_i^M C_i f_i
    \label{eq:sisso_ex}
\end{equation}

The SISSO method constructs features by iteratively applying algebraic operators to primary features in order to construct a more complex space of non-linear features. The number of iterations performed, $N$, is denoted by the rung, $\Phi_N$. The primary features used in this work are the binding site features described in Section~\ref{sec:meth-feats}. 

We use the notation $D_{\Phi_N, M}$ to denote descriptors obtained with dimension $M$ and rung $\Phi_N$. The computational cost of finding the best descriptor increases strongly with the rung and dimension. Here, we considered only dimensions up to 5 and rungs up to 3. For training and validation of rung 3 models we used a subset of the primary features and operators used in feature construction for lower rung models. This was done in order to keep the training time of rung 3 SISSO models practically accessible. Since our main focus here is the interpretability of the SISSO descriptors, which is more easily assessed from the simpler descriptors, we do not pursue more complex models. Further details on the SISSO models used are given in Section~S3.1.3 of the supplementary material.

\subsubsection{WWL-GPR}\label{sec:wwl-gpr}
The Wasserstein Weisfeiler-Lehman (WWL)-GPR model \cite{Xu2022} employs a graph representation of the catalyst surface where the atoms are nodes and the chemical bonds between the atoms are edges in the graph. Here we draw a chemical bond if the distance between two atoms is lower than the sum of their covalent radii, plus \qty{0.2}{\angstrom}.
In this model, the similarity measure input to the GPR model is obtained using a customized version of the WWL graph kernel~\cite{togninalli2019wasserstein,Xu2022}.
The nodes (atoms) of the graph are each decorated with a feature vector. These features include atom-specific features (e.g., Pauli electronegativity) and system-specific features (e.g., work function). The latter are attributed to all nodes corresponding to surface atoms. The information on the pDOS of C, H, and O atoms is excluded, as these atoms do not have a d-band. For missing feature values, a value of 0 is used.

The node features that we assign to each graph node are the same features used in the other (non-graph) models, with one key difference: instead of using averaged geometric descriptors (see features denoted as ``geometric'' and ``stoichiometric'' in Table S1), we use a SOAP descriptor~\cite{bartok2013} for each atom. 
This descriptor is calculated separately for the clean catalyst surface and for the adsorbate in the gas phase.
This choice is motivated by the graph-based representation of the structures, which naturally allows for assigning atom-specific SOAP descriptors directly to the corresponding nodes. Further details are given in Section~S3.1.4 of the supplementary material. We show in Table S5 that the SOAP descriptors slightly increase the predictive performance compared to the ``geometric'' and ``stoichiometric'' features from Table S1.

In contrast, we do not include SOAP features in the non-graph models. This is because maintaining model interpretability is a priority, and averaging SOAP descriptors across binding sites may not yield physically meaningful representations.

\subsubsection{Model Training Procedure}

Adsorption energy models are trained on a dataset of adsorption energies from all systems, with an $80:20$ training--validation and testing split. 
Training--validation and testing splits are stratified, such that 80\% of the datapoints for each material are present in the training--validation set, with the remaining 20\% in the testing dataset. The dataset splits are also stratified with regards to adsorption site type, so that an equal proportion of bridging and on-top configurations are present in each split.
Optimal hyperparameter values are determined by a grid search where mean validation scores from 5-fold cross validation on the training--validation set are used to score each hyperparameter set. The optimized hyperparameters (see Section~S3.1 of the supplementary material) are then used to train a model on the entire training--validation set and the predictions of this model on the test data are then reported as out-of-sample testing predictions.

\section{Results \& Discussion }

\subsection{Dataset}

Table~\ref{tab:dataset} gives an overview of the materials and structures that were considered in the construction of the dataset. The number of stoichiometries and cluster structures for each material within the thresholds described in Section~\ref{sec:meth-ads_e} are shown, along with the number of sites which we enumerate for the clusters of a given material (i.e., the number of binding configurations). Some of the relaxations and evaluations of sampled binding configurations were very slow to converge, in which case we chose to omit them from the dataset. This was particularly an issue with calculations on \ce{Pd}-supported systems, and is the reason we have fewer data points for these.

\begin{table*}
    \centering
    \caption{Materials, and structures therein, used to construct the dataset for training adsorption energy models.}
    \label{tab:dataset}
    \begin{tabular}{lcccc}
        \toprule
         &  \multicolumn{4}{c}{Number of}
         \\\cmidrule{2-5}              
         Material & Stoichiometries & Clusters & Binding configurations & Data points \\\midrule
         \ce{In_yO_x/Au}& 23 & 326 & 4479 & 497  \\
         \ce{In_yO_x/Cu}& 22 & 345 & 5073 &  497 \\
         \ce{In_yO_x/Pd}& 21 & 275 & 4632 &  431 \\
         \ce{Zn_yO_x/Au}& 26 & 153 & 2295 &  496 \\
         \ce{Zn_yO_x/Cu}& 24 & 222 & 3602 &  498 \\
         \ce{Zn_yO_x/Pd}& 25 & 307 & 5036 &  444 \\\midrule
         Total & 141 & 1629 & 25117 &  2863\\\bottomrule
    \end{tabular}
\end{table*}

\begin{figure}
    \centering
    \includegraphics{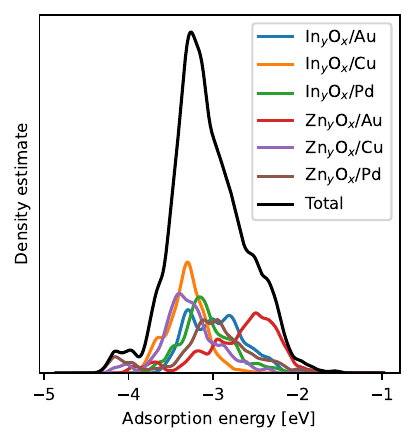}
    \caption{Kernel density estimate plot illustrating the distribution of adsorption energies in the total dataset. Distributions for the individual materials are shown with different colors.}
    \label{fig:be-distribution}
\end{figure}

Figure~\ref{fig:be-distribution} shows the distribution of DFT adsorption energies in the dataset. Statistics on adsorption energy distributions are shown in Table~\ref{tab:be-stats}. Materials with \ce{In_yO_x} clusters have the narrowest distributions. Conversely, materials with \ce{Zn_yO_x} clusters present wider spans of adsorption energies. A trend between the supporting surface and adsorption energy span is noted, with spans increasing in the order \ce{Cu}\,$<$\,\ce{Au}\,$<$\,\ce{Pd}. Accordingly, the widest span of adsorption energies is seen for \ce{Zn_yO_x/Pd(111)} and the narrowest for \ce{In_yO_x/Cu(111)}.
In the assembled data set, about \qty{90}{\percent} of the data points arise from oxygen-rich clusters with a ratio of metal to oxygen atoms in the cluster ($\mathrm{M:O}$) smaller than 3, since these are sampled more often in our data collection procedure.

\begin{table}
    \centering
    \caption{Statistics on adsorption energy distributions for the dataset and individual materials therein.}
    \label{tab:be-stats}
    \begin{tabular}{lS[table-format=-1.2]S[table-format=-1.2]S[table-format=1.2]}
    \toprule
    & \multicolumn{3}{c}{$E_{\mathrm{ads}}$ [\unit{\eV}]}
    \\\cmidrule{2-4}
    Material & {Min.} & {Max.} & {Span} \\\midrule
    \ce{In_yO_x/Au} & -3.62 & -1.86 & 1.76 \\
    \ce{In_yO_x/Cu} & -3.84 & -2.52 & 1.32 \\
    \ce{In_yO_x/Pd} & -3.77 & -1.75 & 2.02 \\
    \ce{Zn_yO_x/Au} & -4.02 & -1.55 & 2.47 \\
    \ce{Zn_yO_x/Cu} & -4.19 & -2.02 & 2.17 \\
    \ce{Zn_yO_x/Pd} & -4.31 & -1.54 & 2.77 \\\midrule
    Total & -4.31 & -1.54 & 2.77 \\\bottomrule
    \end{tabular}
\end{table}

\subsection{Adsorption Energy Predictions}

\begin{figure*}
    \centering
    \includegraphics{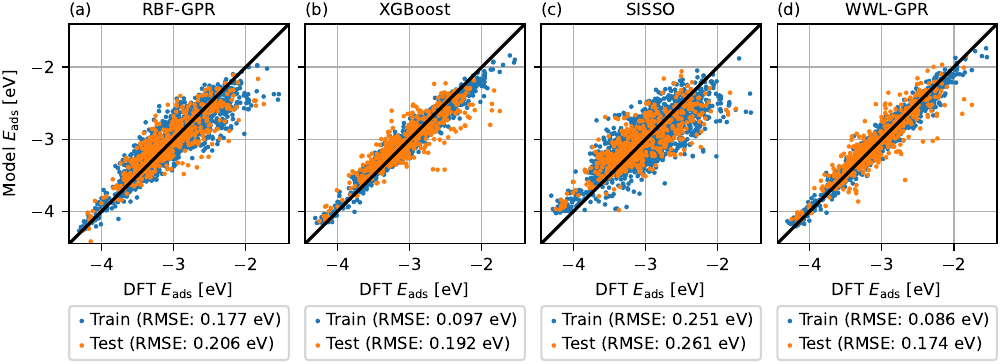}
    \caption{Parity plots for adsorption energy models.}
    \label{fig:parityplots}
\end{figure*}

\begin{table*}
    \centering
    \caption{Root mean square error (RMSE), mean absolute error (MAE), and maximum absolute error (MaxAE) for adsorption energy models trained on the total training--validation dataset. Errors provided are for predictions on test splits of individual materials. Boldface denotes the lowest error for each dataset.}
    \label{tab:model-errors}
    \begin{tabular}{@{}cS[table-format=1.3]S[table-format=1.3]S[table-format=1.3]S[table-format=1.3]S[table-format=1.3]S[table-format=1.3]S[table-format=1.3]S[table-format=1.3]S[table-format=1.3]S[table-format=1.3]S[table-format=1.3]S[table-format=1.3]@{}}
        \toprule
        & \multicolumn{4}{c}{RMSE [\unit{\eV}]} & \multicolumn{4}{c}{MAE [\unit{\eV}]} & \multicolumn{4}{c}{MaxAE [\unit{\eV}]} \\\cmidrule(lr){2-5} \cmidrule(lr){6-9} \cmidrule(lr){10-13} 
        & {GPR\textsuperscript{RBF}} & {XGBoost} & {SISSO} & {GPR\textsuperscript{WWL}} & {GPR\textsuperscript{RBF}} & {XGBoost} & {SISSO} & {GPR\textsuperscript{WWL}} & {GPR\textsuperscript{RBF}} & {XGBoost} & {SISSO} & {GPR\textsuperscript{WWL}} \\\midrule
        \ce{In_yO_x/Au} & 0.210 & 0.206          & 0.257 & \textbf{0.136} & 0.146 & 0.135          & 0.205 & \textbf{0.098} & 0.728          & 0.906          & 0.800 & \textbf{0.587} \\
        \ce{In_yO_x/Cu} & 0.139 & \textbf{0.118} & 0.182 & 0.128          & 0.108 & \textbf{0.092} & 0.143 & 0.096          & 0.386          & \textbf{0.343} & 0.517 & 0.346          \\
        \ce{In_yO_x/Pd} & 0.246 & 0.237          & 0.289 & \textbf{0.236} & 0.177 & 0.159          & 0.222 & \textbf{0.153} & \textbf{0.790} & 0.918          & 0.873 & 0.890          \\
        \ce{Zn_yO_x/Au} & 0.206 & 0.181          & 0.263 & \textbf{0.167} & 0.163 & 0.136          & 0.211 & \textbf{0.125} & 0.639          & 0.500          & 0.625 & \textbf{0.445} \\
        \ce{Zn_yO_x/Cu} & 0.195 & 0.176          & 0.285 & \textbf{0.170} & 0.140 & 0.135          & 0.217 & \textbf{0.128} & 0.632          & \textbf{0.533} & 0.869 & 0.538          \\
        \ce{Zn_yO_x/Pd} & 0.231 & 0.222          & 0.282 & \textbf{0.198} & 0.173 & 0.154          & 0.227 & \textbf{0.140} & \textbf{0.657} & 0.740          & 0.766 & 0.762          \\
        \midrule
        Total & 0.206 & 0.192 & 0.261 & \textbf{0.174} & 0.150 & 0.134 & 0.203 & \textbf{0.122} & \textbf{0.639} & 0.918 & 0.873 & 0.890 \\\bottomrule
    \end{tabular}
\end{table*}

Parity plots illustrating the predictive accuracy of adsorption energy models trained on the total dataset are shown in Figure~\ref{fig:parityplots}. Results for the SISSO model are drawn from predictions using a $D_{3,5}$ descriptor. Values for the RMSE of predicted adsorption energies on training--validation and testing splits are indicated for each model. Based on these RMSE values, the WWL-GPR model provides the most accurate predictions, closely followed by the XGBoost model. Similar to previous results~\cite{Xu2022}, the improved accuracy of WWL-GPR can be attributed to the graph representation employed in this model, which is particularly beneficial for more complex adsorbates such as formate that may occur in both mono- and bidentate binding configurations. Although the RBF-GPR model yields higher RMSE values on both splits, the difference between training and testing RMSEs is smaller than for the WWL-GPR and XGBoost models. This is likely due to some overfitting occuring in the XGBoost and WWL-GPR models. The SISSO model provides the least accurate adsorption energy predictions. As noted in Section~\ref{sec:sisso}, it is possible that a marginally better predictive accuracy could be obtained with more complex models of higher rung and dimension, albeit at a much increased computational cost of the descriptor identification.

RMSE values for the adsorption energy models on testing splits of individual materials are provided in Table~\ref{tab:model-errors}, along with alternative error measures. From this we again see that the WWL-GPR model provides the most accurate adsorption energy predictions, scoring the lowest RMSEs and MAEs across all materials, except for \ce{In_yO_x/Cu}, where XGBoost is slightly better. The RBF-GPR model sometimes yields lower maximum absolute errors (MaxAE) than WWL-GPR and XGBoost. This may also be attributed to overfitting in the WWL-GPR and XGBoost models, yielding outliers in the testing predictions.

Comparison of training and testing RMSE values for each of the models reveals a familiar compromise in machine learning between model complexity and predictive accuracy. More complex models offer greater predictive accuracy at the cost of some overfitting, as revealed by the difference in RMSE of XGBoost predictions on training and test data. Conversely, simpler models are less prone to overfitting, at the expense of predictive accuracy. This is most clearly observed with the SISSO model, yielding the highest RMSE predictions on training and testing data, while the difference between the two RMSEs is the smallest. 

It is worth highlighting that the prediction targets for all models tested here are adsorption energies drawn from relaxed cluster\,+\,formate configurations, while the features informing these predictions are obtained from the clean clusters. It is in directly bridging this gap from the clean cluster structure to the relaxed adsorption energy, along with their low computational cost, that these models offer utility as a tool in catalyst screening. Within this context, the models allow for cheap prediction of adsorption energies at sites near the cluster--surface interface. In case of the RBF-GPR, WWL-GPR and XGBoost models, these predicted adsorption energies for unseen data points are shown in Figure~\ref{fig:parityplots} to be within an error corresponding to approximately \qty{7}{\percent} of the total span of adsorption energies. Such an error is sufficiently low as to allow for identification of sites within some favorable range of adsorption energies. This quick identification then allows for a directed investigation of sites with more expensive methods if more accurate adsorption energy values are required. 
Furthermore, as we show in Section~S4 of the supplementary material, the prediction errors of all three models are found to be significantly lower than corresponding errors in adsorption energies calculated using the foundational pre-trained ML interatomic potentials CHGNet~\cite{deng_2023_chgnet} and MACE-MP-0a~\cite{batatia2023foundation}.
The main limitation of the models developed here is that they cannot be expected to work well in cases where the adsorbate dissociates or the surface reconstructs significantly during the adsorption process, since in contrast to ML interatomic potentials, they do not model the entire potential energy surface of the system. We have previously shown that this limits the effectiveness of these models for amorphous nanosilicate clusters, where adsorption processes may result in bond breaking and cluster restructuring~\cite{Andersen2023}. However, for the supported oxide nanoclusters considered in this work, cluster reconstruction seems to be a lesser issue.

\begin{figure*}
    \centering
    \includegraphics{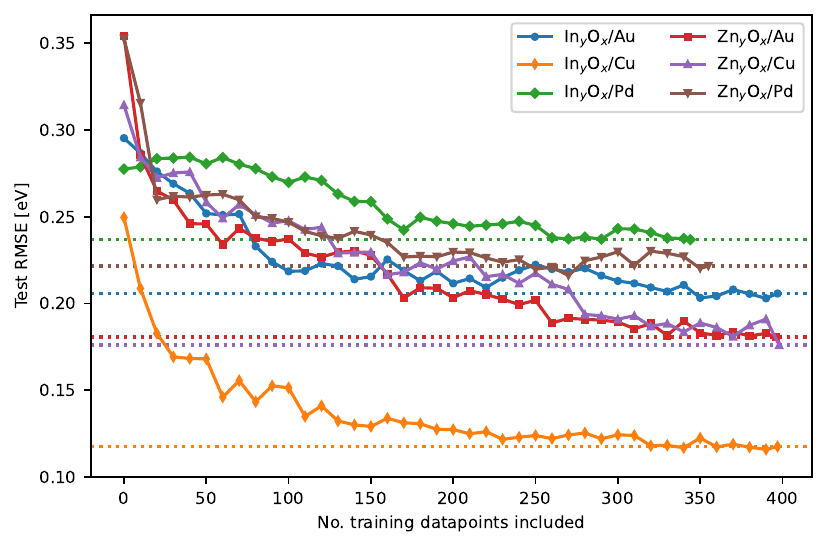}
    \caption{Learning curve for XGBoost models trained on incrementally more training data from a material, which is initially left out of the total training data set. The RMSE reported is for test data of the omitted material.}
    \label{fig:learningcurve}
\end{figure*}

When applying the models to an unseen material, i.e., a new combination of metal oxide cluster and metal substrate, it will be necessary to acquire new training data in order to obtain accurate predictions. Predictions of the XGBoost model on testing data from a material that was entirely or partially left out of the training data are shown in Figure~\ref{fig:learningcurve}. Training data points from the unseen material are gradually included in accordance with the FPS procedure outlined in Section~\ref{sec:meth-bindingsites}. RMSE values from predictions on the test data of materials, which are completely absent in the training data, are on the order of \qty{0.1}{\eV} worse than when the full set of training points is included. The degree of improvement in prediction accuracy when incorporating more training data varies between materials, but in most cases the most drastic improvement is seen within the inclusion of the first \num{200} data points. Exceptions to this are \ce{Zn_yO_x/Au(111)} and \ce{Zn_yO_x/Cu(111)} where the RMSEs of test predictions only approach their final values after large dips at around \num{250} data points. Across all materials the inclusion of \num{280} data points during training is sufficient to obtain an RMSE which remains within \qty{0.02}{\eV} of that obtained using the full training dataset.

A similar investigation was conducted for the RBF-GPR model, but with only up to 100 training points included for each material. Here we found that the inclusion of 100 data points from the unseen material was enough to obtain an RMSE on the test split converged within \qty{0.06}{\eV} across all materials. For almost all materials this threshold is met with fewer added data points by the XGBoost model, except for \ce{Zn_yO_x/Cu(111)}, which required 130 data points. Thus, XGBoost may offer advantages in terms of data efficiency. Among all the models tested, it is also the model with the lowest computational cost of model training, as outlined in Table S6 in the supplementary material. For prediction, the analytical SISSO model has the lowest computational cost, closely followed by XGBoost.

\subsection{Model Interpretability}

Another useful aspect of the XGBoost model is that it allows for model interpretability through the feature importance scores, from which we provide the six highest scoring features in Figure~\ref{fig:xgb-fi} (the full list is given in Figure~S7 in the supplementary information). The two highest-scoring features are the work function of the surface ($\phi$) and the Pauling electronegativity of the binding site atoms ($\chi$). Since the importance scores reported by the model are normalized, these two features account for approximately \qty{36}{\percent} of the gain in the model when used in a split. The emergence of the work function as the highest-scoring feature is reasonable, given that the Pearson correlation coefficient between it and the adsorption energies is \num{0.62}, which is the highest among all features. This is also evident when inspecting the $D_{0,5}$ descriptor identified by SISSO, as shown in Equation~\eqref{eq:r0d5_desc}. Here we see the work function identified as the first term in the SISSO descriptor, with d- and sp-band features appearing in subsequent terms. The work function also appears in the first terms of the higher-rung descriptors (Equations~\eqref{eq:r1d5_desc}--\eqref{eq:r3d5_desc_simp}). Prominence of the work function in the initial terms of each SISSO descriptor suggests that it, along with higher-rung features incorporating it, serves as an effective predictor of the adsorption energy.

The correlation between the work function and adsorption energies, as highlighted by the SISSO descriptors and XGBoost importance scores, further connects to the stoichiometric properties of the clusters. The oxygen content of clusters correlates with the work function, with oxygen-poor structures exhibiting lower values (see Figure~S8 in the supplementary information). Figure~\ref{fig:cmo_grid_be} illustrates how this translates to a relationship between the stoichiometry of clusters and the mean adsorption energy at sites on those clusters. There is a clear trend that clusters with lower oxygen contents have binding sites which exhibit lower mean adsorption energies than in more oxygen-rich clusters. The trend is likely caused by the presence of metal atoms that do not coordinate to any oxygen in oxygen-poor clusters, which are more reactive than their oxygen-coordinated counterparts. This observation aligns well with previous literature findings on the importance of oxygen vacancies in bulk oxides for the formation and stabilization of formate and for the \ce{CO2} hydrogenation reaction in general~\cite{wang2017,frei2018,chen2021,cheula2024}.

A key difference in the interpretation of the two models lies in the significance of the Pauling electronegativity of the binding site atoms. The following comparative analysis omits the rung 3 SISSO descriptor, as it is trained with a limited set of primary features selected as the 11 highest scoring features based on the XGBoost feature importance metric. Consequently, this complicates the analysis of primary feature importance by examining their frequency in rung 3 descriptors. 
The feature importance score reported by XGBoost points to the Pauling electronegativity ($\chi$) being a significant predictor of the adsorption energy. On the other hand, the absence of $\chi$ from all but one term of the SISSO descriptors suggests that it is less important than d- and sp-band features, which appear more prominently. This discrepancy is likely due to the different nature of the two models and the way they identify trends in the dataset. 
The work function is closely tied to cluster oxygen content, and $\chi$ essentially codes for the nature of the atomic elements present in the binding site. Splitting the data along these features can then roughly group data points by material, binding site type, and oxygen content of the cluster. This in turn may explain the high importance scores for both of these features. In contrast, the SISSO models learn adsorption energies by identifying linear fits in a space of constructed features. Here, the discrete distribution of $\chi$ values may be less favorable than the more continuous spread of other features, such as the d- and sp-band features.

The d-band center ($\epsilon_\mathrm{d}$), width ($W_\mathrm{d}$), and filling ($f_\mathrm{d}$) each appear in several of the descriptors identified by SISSO. The d-band width also has the third highest feature importance score reported by the XGBoost model. The prominence of d-band features in both models seems reasonable in light of the d-band model formulated by \citeauthor{hammer1995noble}, which highlights the importance of the d-band position as a predictor of adsorption energies on transition metals~\cite{hammer1995noble}.  

\begin{figure}
    \centering
    \includegraphics{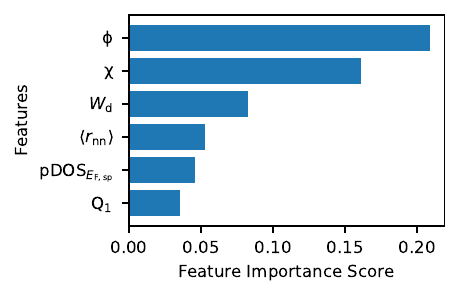}
    \caption{Features with feature importance scores above 0.035 from the XGBoost model trained on the total dataset.}
    \label{fig:xgb-fi}
\end{figure}

\begin{equation}
    D_{0,5} = 
    \begin{bmatrix}
        \phi \\
        W_{\mathrm{d}} \\
        \epsilon_{\mathrm{sp}} \\
        \epsilon_{\mathrm{d}} \\
        f_{\mathrm{d}}
    \end{bmatrix}
    \label{eq:r0d5_desc}
\end{equation}

\begin{equation}
    D_{1,5} = 
    \begin{bmatrix}
        \phi \cdot f_{\mathrm{d}} \\
        q_{\mathrm{B}} / Q_1 \\
        Q_1 + f_{\mathrm{sp}} \\
        \log \mathrm{DOS}_{E_{\mathrm{F}}} \\
        \chi + f_{\mathrm{d}}  \\
    \end{bmatrix}
    \label{eq:r1d5_desc}
\end{equation}

\begin{equation}
    D_{2,5} = 
    \begin{bmatrix}
        (\sigma_{\mathrm{d}} - \phi) \cdot (\phi \cdot W_{\mathrm{d}}) \\
        | (\phi \cdot \epsilon_{\mathrm{sp}}) - (\mu_d \cdot W_{\mathrm{sp}}) | \\
        (\phi + \epsilon_{sp}) \cdot (\mu_{\mathrm{d}} \cdot f_{\mathrm{sp}}) \\
        | (f_{\mathrm{d}} / Q_1) - (\phi / W_{\mathrm{d}}) | \\
        \log\mathrm{pDOS}_{E_{\mathrm{F}}} \cdot (\mathrm{pDOS}_{E_{\mathrm{F}}} / W_{\mathrm{sp}}) 
    \end{bmatrix}
    \label{eq:r2d5_desc}
\end{equation}

\begin{equation}
    D_{3,5} = 
    \begin{bmatrix}
        \dfrac{\phi^4 \cdot W_{\mathrm{d}}}{Q_1 \cdot \chi \cdot f_{\mathrm{d}}} \\[1.1em]
        \left| \dfrac{\log\mathrm{pDOS}_{\mathrm{sp}, E_{\mathrm{F}}}\cdot\phi}{\epsilon_{\mathrm{d}}} - \dfrac{\log f_{\mathrm{d}}\cdot\epsilon_{\mathrm{d}}}{\epsilon_{\mathrm{sp}}} \right| \\[1.1em]
        \left| \dfrac{\chi\cdot\epsilon_{\mathrm{d}}}{\chi + f_{\mathrm{d}}} - \dfrac{\epsilon_{\mathrm{d}}-\epsilon_{\mathrm{sp}}}{\log\langle r_{\mathrm{nn}} \rangle} \right| \\[1.1em] 
        \left| \dfrac{\chi\cdot\phi}{Q_1^2} - \dfrac{\epsilon_{\mathrm{sp}}^2}{Q_1 \cdot W_d} \right| \\[1.1em]
        \left| \dfrac{f_{\mathrm{d}}-\chi}{W_{\mathrm{d}}} - \dfrac{f_{\mathrm{d}} \cdot \log W_{\mathrm{d}}}{\phi} \right|
    \end{bmatrix}
    \label{eq:r3d5_desc_simp}
\end{equation}

\begin{figure*}
    \centering
    \includegraphics{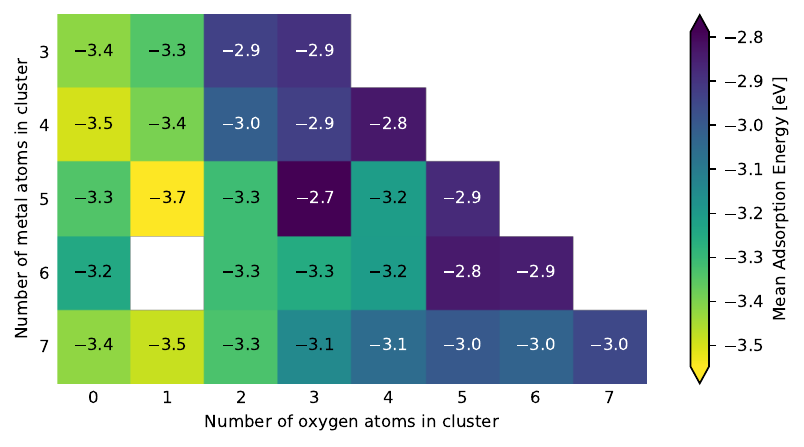}
    \caption{Grid plot illustrating the relationship between the oxygen content of clusters and the mean adsorption energy at sites on those clusters. \ce{M6O} stoichiometries are excluded in accordance with our previous \textit{ab initio} thermodynamics study \cite{kempen2025inverse}.}
    \label{fig:cmo_grid_be}
\end{figure*}

\section{Conclusion}

In summary, this work presents an investigation of formate adsorption energies at the metal-oxide interface of inverse catalyst materials, specifically \ce{In_yO_x} and \ce{Zn_yO_x} clusters supported on \ce{Au(111)}, \ce{Cu(111)} and \ce{Pd(111)}, using DFT as well as ML models with varying degrees of interpretability. We implement a workflow for enumerating binding sites at the metal-oxide interface, and sample a subset of these sites from each material to evaluate with DFT.  Features in the dataset are drawn from the clean nanocluster structures to capture stoichiometric, geometric, and electronic properties of both the binding site atoms and the full nanocluster system. The predictive performances of RBF-GPR, XGBoost, SISSO and WWL-GPR models trained on this dataset are evaluated, with the graph-based WWL-GPR model yielding the most accurate adsorption energy predictions on the test data across all materials, closely followed by XGBoost and RBF-GPR. The number of sampled training data points required to approach the prediction error of the full model has been tested for XGBoost and was found to vary between materials. Predictions on test data yield an RMSE within \qty{0.02}{\electronvolt} of the full model once \num{90} data points are included for \ce{In_yO_x/Au(111)}, while \num{280} data points are required for \ce{Zn_yO_x/Cu(111)}.

Analysis of descriptors identified by SISSO and feature importance scores reported by XGBoost points to the work function as a strong predictor of adsorption energies. The work function is highly correlated with the oxygen content of clusters, leading to a trend where sites on clusters with lower oxygen content tend to exhibit lower adsorption energies. From a catalyst design perspective this suggests that the oxygen content of clusters can be viewed as a central tuning parameter to control the expected adsorption energies of formate, and thereby likely also the activity of the inverse catalyst for \ce{CO2} hydrogenation.

As we show in our previous work \cite{kempen2025inverse}, the most thermodynamically stable composition and structure of the oxide clusters can be tuned either by the oxygen chemical potential (i.e., the temperature and pressure of gas-phase species under reaction conditions) or by the nature of the metal substrate. In particular, reduced clusters tend to be stabilized under the reducing conditions typical for \ce{CO2} hydrogenation. And at a given oxygen chemical potential, the susceptibility to oxidation of the metal oxide cluster is highly dependent on the metal substrate and increases in the sequence Au\,$<$\,Cu\,$<$\,Pd. In this respect, the choice of metal substrate will of course also further tune the adsorption energies at the metal-oxide interface through other important features of the binding site atoms such as the Pauling electronegativity and features related to the pDOS.

Finally, we note that the developed models predict adsorption energies of relaxed configurations at all sites of a cluster, while only being informed by features of the clean cluster. Bridging this gap allows for significant savings in computational resources. We expect that the workflow can readily be extended to other adsorbates, allowing for the simultaneous prediction of many combinations of sites/adsorbates at the cost of a single DFT calculation of the clean cluster. The extension to other metal-oxide combinations is also expected to be possible with a moderate investment of additional DFT training data calculations, as illustrated by our learning curve analysis. Thus, the developed approach can serve to accelerate catalyst screening by allowing for rapid identification of promising active sites and prediction of parameters required for microkinetic modeling.

\section*{Supplementary Material}

See the supplementary material for further details on the used features and binding energy models as well as a comparison to results obtained using the  foundational pre-trained models CHGNet and MACE-MP-0a.

\begin{acknowledgments}
    M.A. and R.C. acknowledge funding from the European Union’s Horizon 2020 research and innovation programme under the Marie Skłodowska-Curie grant agreement No 754513 and 101108769, respectively. M.A. acknowledges funding from The Aarhus University Research Foundation, the Danish National Research Foundation through the Center of Excellence `InterCat' (grant no.\ DNRF150), and VILLUM FONDEN (grant no.\ 37381). Computational support was provided by the Centre for Scientific Computing Aarhus (CSCAA) at Aarhus University.
\end{acknowledgments}

\section*{Author Declarations}

\subsection*{Conflict of Interest}

The authors have no conflicts to disclose.

\subsection*{Author Contributions}

\textbf{Marius Juul Nielsen}: Conceptualization (equal); Investigation (lead); Methodology (lead); Software (lead); Visualization (lead); Writing -- original draft (lead); Writing -- review \& editing (equal).

\textbf{Luuk H. E. Kempen}: Conceptualization (supporting); Investigation (supporting); Methodology (supporting); Writing -- review \& editing (equal); Supervision (supporting).

\textbf{Julie de Neergaard Ravn}: Investigation (supporting); Software (supporting).

\textbf{Raffaele Cheula}: Investigation (supporting); Writing -- review \& editing (supporting); Supervision (supporting).

\textbf{Mie Andersen}: Conceptualization (equal); Investigation (supporting); Writing -- review \& editing (equal); Supervision (lead); Resources (lead); Funding acquisition (lead).

\section*{Data Availability}

The data that support the findings of this study will be made openly available via Zenodo upon publication.

\section*{References}
\bibliography{references}

\end{document}


\maketitle

\newpage

\section{Data acquisition and features}

Table \ref{tab:feats} shows all of the 27 features included in the dataset. In the case of sites with two binding site atoms, we calculate values for site specific features by taking the mean of the feature value for the two atoms. 

\begin{figure}[H]
    \centering
    
    \includegraphics{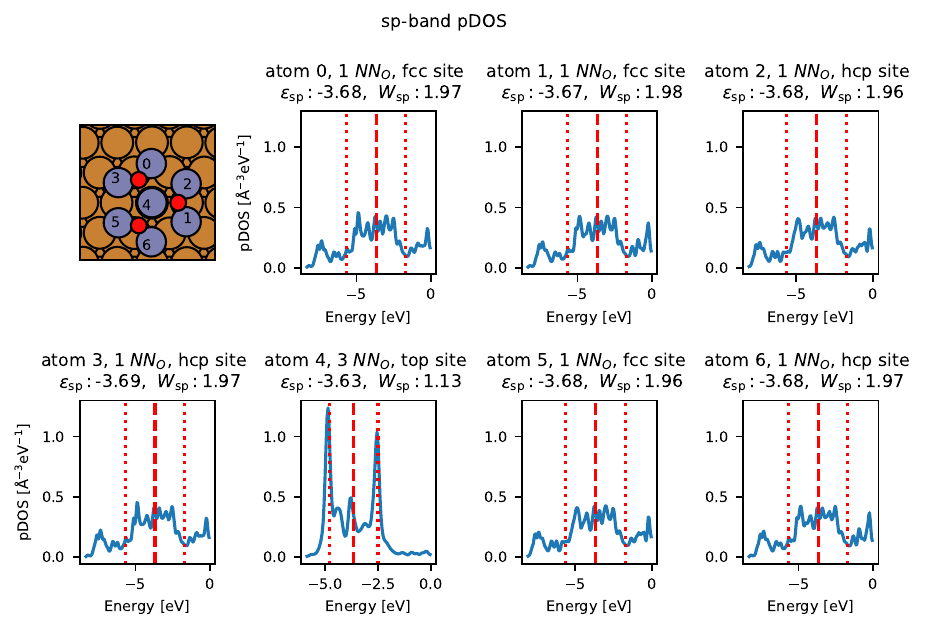}
    \caption{Example of sp-band centers and widths calculated for \ce{Zn} atoms.}
    \label{fig:zn-pdos-sp}
\end{figure}

\begin{figure}[H]
    \centering
    \includegraphics{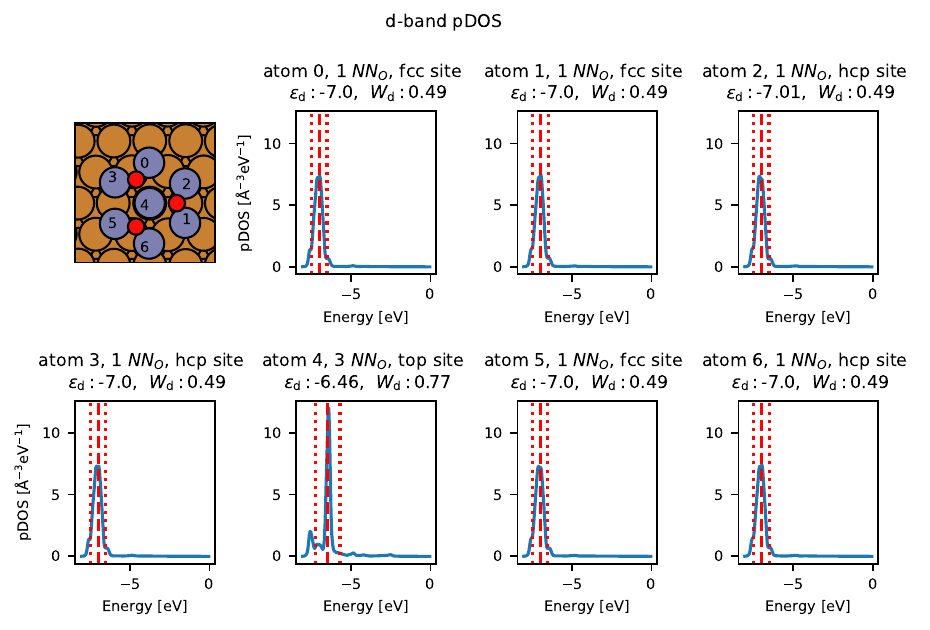}
    \caption{Example of d-band centers and widths calculated for \ce{Zn} atoms.}
    \label{fig:zn-pdos-d}
\end{figure}

\begin{figure}[H]
    \centering
    \includegraphics{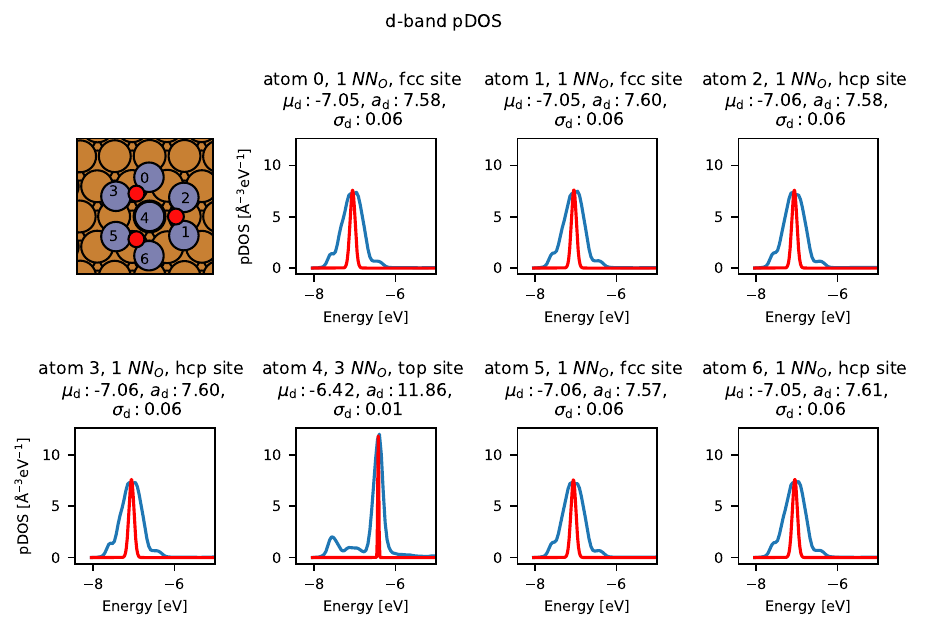}
    \caption{Example of Gaussian fits on peaks in the d-band pDOS for \ce{Zn} atoms. A restricted range of energies are shown to make the widths of the Gaussians more clear}
    \label{fig:zn-pdos-d-gauss}
\end{figure}

\begin{figure}[H]
    \centering
    \includegraphics{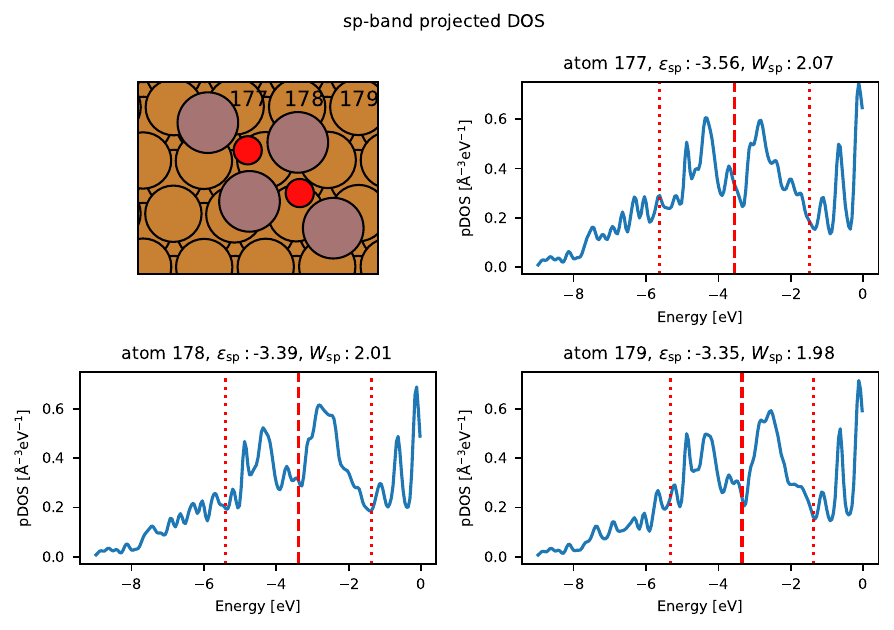}
    \caption{Example of sp-band centers and widths calculated for \ce{Zn} atoms.}
    \label{fig:cu-pdos-sp}
\end{figure}

\begin{figure}[H]
    \centering
    \includegraphics{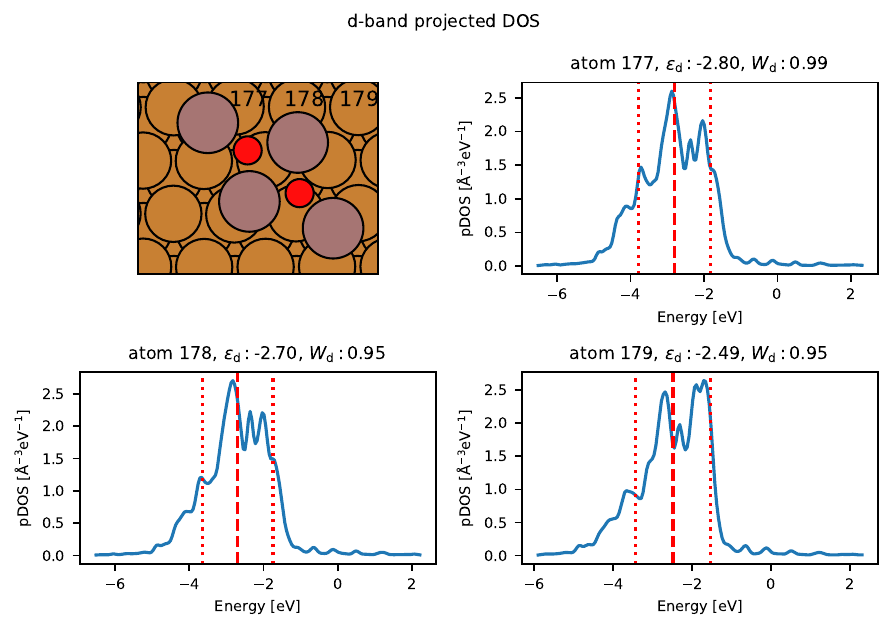}
    \caption{Example of d-band centers and widths calculated for \ce{Zn} atoms.}
    \label{fig:cu-pdos-d}
\end{figure}

\begin{figure}[H]
    \centering
    \includegraphics{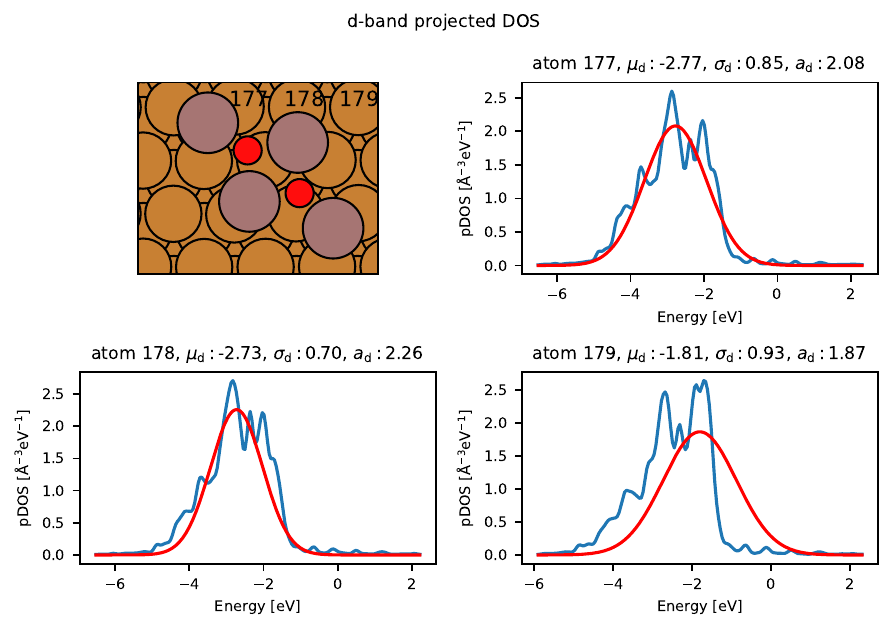}
    \caption{Example of Gaussian fits on peaks in the d-band pDOS for \ce{Cu} atoms.}
    \label{fig:cu-pdos-d-gauss}
\end{figure}

When calculating features using pDOS data extracted from the clean clusters we chose to limit the range of energies considered. For d-band pDOS this limited range is defined as being between the lowest and highest energies with a pDOS value of \qty{0.01}{\angstrom^{-3}\eV^{-1}}, from the energy where the pDOS is maximum. If the upper limit as defined in this way lies below the Fermi level, we change the upper limit to be the Fermi level. The limits for sp-band pDOS are similarly defined, but where the upper limit is constrained to at most being the Fermi level. Examples of calculated sp- and d-band centers are shown in Figures \ref{fig:zn-pdos-sp}, \ref{fig:zn-pdos-d} and \ref{fig:zn-pdos-d-gauss} for \ce{Zn} atoms of a \ce{Zn7O3} cluster. Similiar plots are included for \ce{Cu} atoms near an \ce{In4O2} cluster in Figures \ref{fig:cu-pdos-sp}, \ref{fig:cu-pdos-d} and \ref{fig:cu-pdos-d-gauss}.

\begin{table}[H]
\begin{tabularx}{\textwidth}{cXl}
\toprule
Symbol & Feature description & Type         \\ \midrule
$S$               & Number of binding site atoms  & Stoichiometric, Site \\
$\mathrm{CM:O}$   & Ratio of metal to oxygen atoms in the cluster & Stoichiometric, Global   \\ 
\addlinespace[0.5em]
$\mathrm{nn_O}$   & Number of nearest-neighbor oxygen atoms to the binding site  & Stoichiometric, Site     \\ 
\addlinespace[0.5em]
$\mathrm{nn_{CM}}$  & Number of nearest-neighbor cluster metal atoms to the binding site & Stoichiometric, Site     \\ 
\addlinespace[0.5em]
$\langle r_{\mathrm{nn}} \rangle$  & Mean distance to nearest neighbor atoms at the binding site  & Geometric, Site \\ 
\addlinespace[0.5em]
$Q_l$ & Angular-resolved order parameters for the quantum number $l \in \{1, \ldots, 6\}$ centered at the binding site atom(s) & Geometric, Site \\
\addlinespace[0.5em]
$\phi$ & Work function of the clean cluster structure & Electronic, Global    \\ 
\addlinespace[0.5em]
$\mathrm{DOS}_{E_{\mathrm{F}}}$  & Density of states at the Fermi energy for the clean cluster structure & Electronic, Global \\ 
\addlinespace[0.5em]
$\mathrm{pDOS}_{E_{\mathrm{F}}}$ & Projected density of states at the Fermi energy for the binding site atoms & Electronic, Site  \\
\addlinespace[0.5em]
$\mathrm{pDOS}_{\mathrm{sp}, E_{\mathrm{F}}}$ & sp-band projected density of states at the Fermi energy & Electronic, Site \\ 
\addlinespace[0.5em]
$\mathrm{pDOS}_{\mathrm{d}, E_{\mathrm{F}}}$  & d-band projected density of states at the Fermi energy  & Electronic, Site \\ 
\addlinespace[0.5em]
$\mathrm{\chi}$       & Tabulated values for Pauli electronegativity for site atoms  & Electronic, Site    \\ 
\addlinespace[0.5em]
$q_{\mathrm{B}}$ & Bader charge of binding site atoms  & Electronic, Site    \\ 
\addlinespace[0.5em]
$\epsilon_{\mathrm{d}}$  & Center of the site-projected d-band & Electronic, Site  \\ 
\addlinespace[0.5em]
$W_{\mathrm{d}}$  & Width of the site-projected d-band  & Electronic, Site   \\ 
\addlinespace[0.5em]
$f_{\mathrm{d}}$      & Filling of the d-band   & Electronic, Site    \\ 
\addlinespace[0.5em]
$\mu_{\mathrm{d}}$  & Center of the Gaussian fitted to the pDOS of the d-band  & Electronic, Site     \\ 
\addlinespace[0.5em]
$\sigma_{\mathrm{d}}$ & Width of the Gaussian fitted to the pDOS of the d-band  & Electronic, Site    \\ 
\addlinespace[0.5em]
$a_{\mathrm{d}}$      & Height of the Gaussian fitted to the pDOS of the d-band   & Electronic, Site     \\ 
\addlinespace[0.5em]
$\epsilon_{\mathrm{sp}}$ & Center of the site-projected sp-band & Electronic, Site  \\
\addlinespace[0.5em]
$W_{\mathrm{sp}}$ & Width of the site-projected sp-band & Electronic, Site  \\ 
\addlinespace[0.5em]
$f_{\mathrm{sp}}$          & Filling of the sp-band  & Electronic, Site   \\ 
\bottomrule 
\end{tabularx}
\label{tab:feats}
\caption{Table of features included in the dataset}
\end{table}

\clearpage

\subsection{Feature Importances}

\begin{figure}[H]
    \centering
    \includegraphics{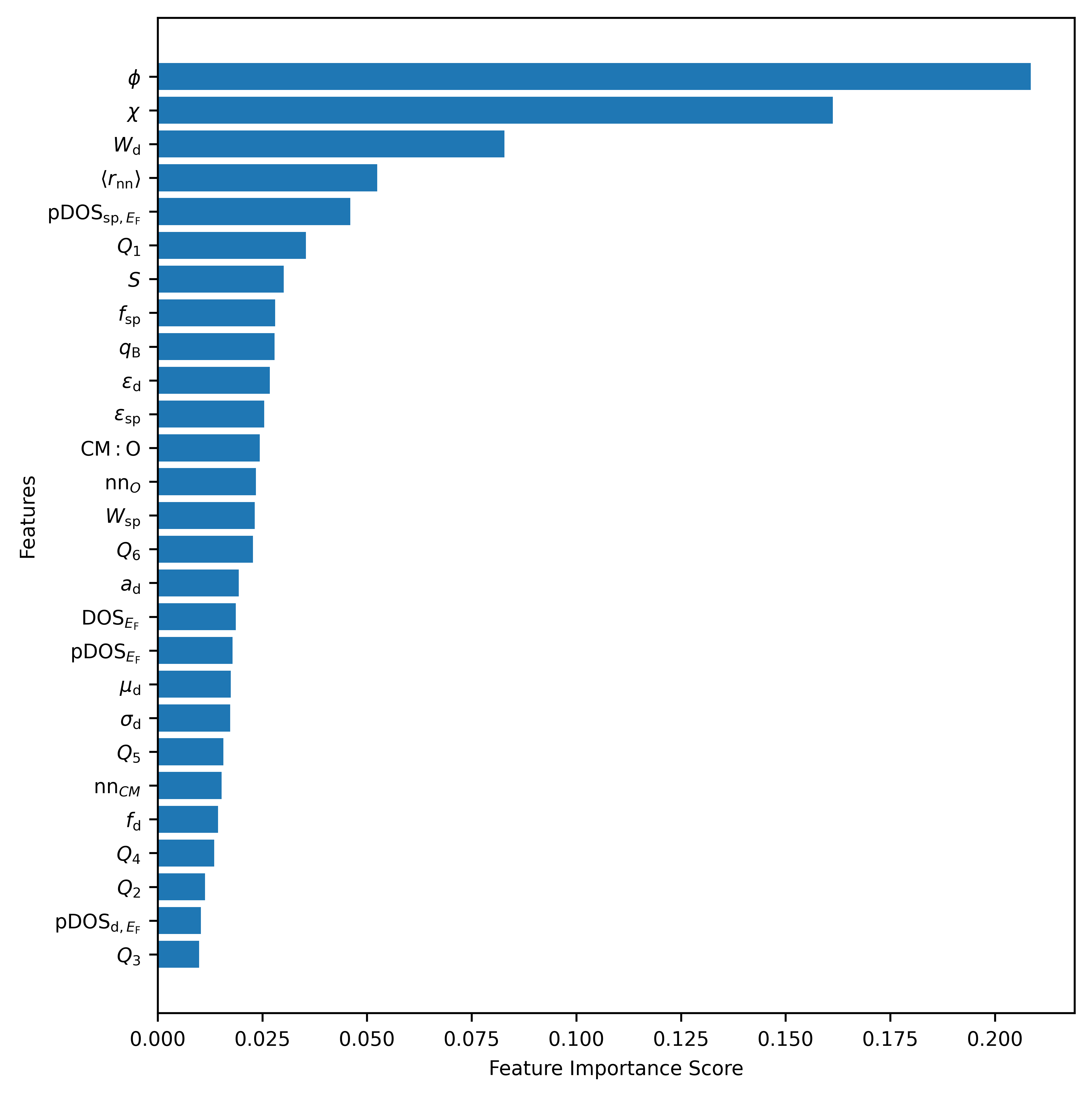}
    \caption{Feature importance scores for all features within training data}
    \label{fig:fi_all}
\end{figure}

Figure \ref{fig:fi_all} shows feature importance scores for all features in the dataset, as reported by the XGBoost model trained on the entire training--validation dataset.

\subsection{SOAP parameters}

To ensure diverse representation of local environments when choosing which formate binding configurations to evaluate with DFT, we use farthest point sampling (FPS) in SOAP space. We calculate the SOAP descriptor with the Dscribe package \cite{himanen2020dscribe}. The SOAP parameters \texttt{r\_cut}, \texttt{n\_max}, \texttt{l\_max}, and \texttt{sigma} are set to 5, 3, 2, and 1, respectively, and a polynomial function \cite{caro2019optimizing} is used for the weighting of the atomic density, with parameters \texttt{r0}, \texttt{m}, and \texttt{c} set to 5, 2, and 1, respectively.

\clearpage

\section{Dataset Trends}

\begin{figure}[H]
    \centering
    \includegraphics[width=\linewidth]{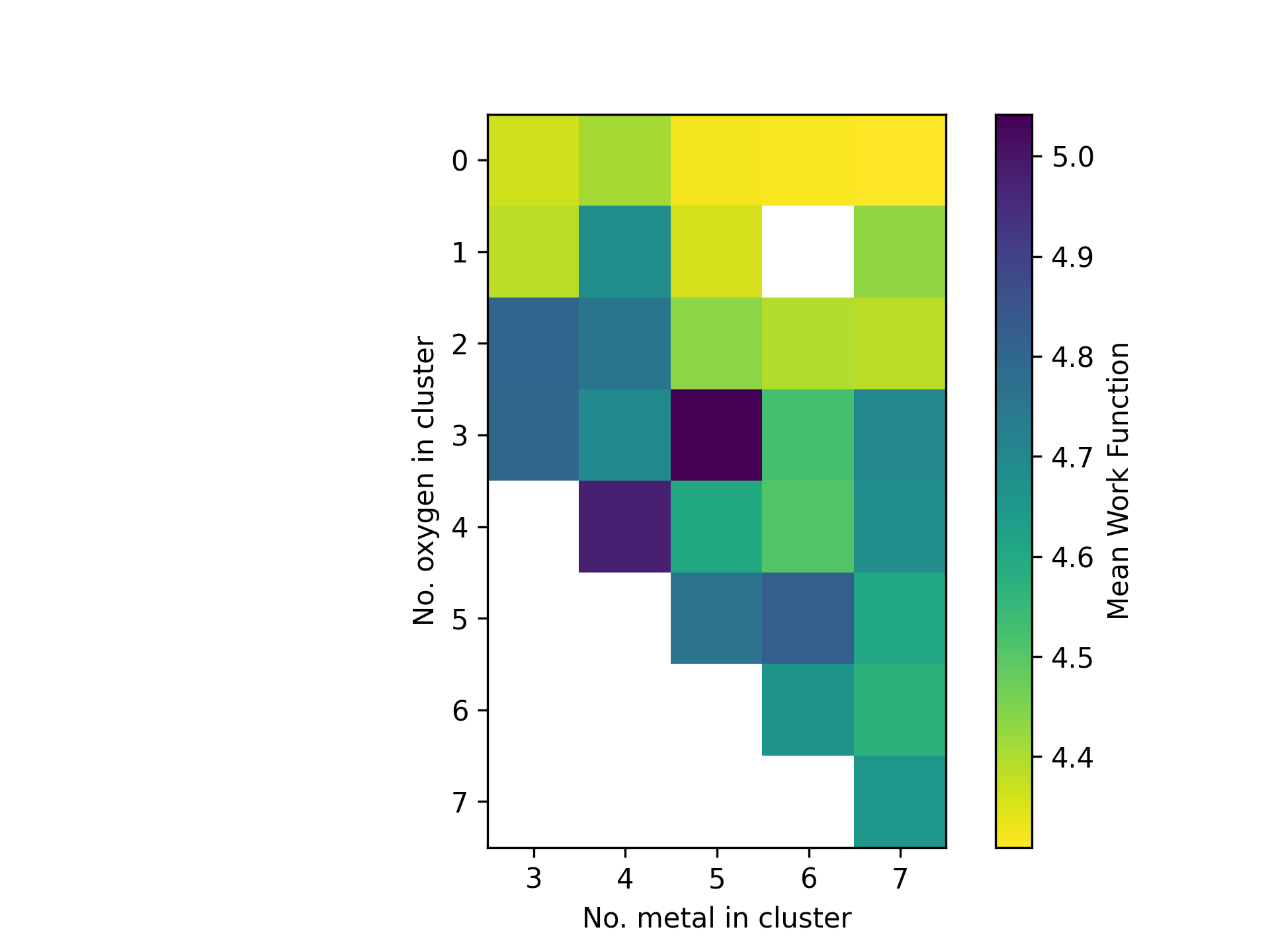}
    \caption{Grid plot illustrating correlation between oxygen content of clusters and work function}
    \label{fig:cmogrid-be}
\end{figure}

Here we include Figure \ref{fig:cmogrid-be} which illustrates the correlation between cluster oxygen content and the work function of the catalyst surface. 

\clearpage

\section{Binding energy models}

\subsection{Model hyperparameters}
Here we show the optimal hyperparameters for each binding energy model type, as determined via a grid search with 5-fold cross validation on the training--validation set.

\subsubsection{RBF-GPR}

We use the scikit-learn \texttt{GaussianProcessRegressor} implementation with a \texttt{Constant}*\texttt{RBF} + \texttt{White} kernel,
\begin{equation}
    k(x_i, x_j) = A^2 \exp\left(-\frac{1}{2l^2}(x_i - x_j)^2\right) + \sigma_n^2\delta_{i,j}
\end{equation}

Optimized values of the kernel hyperparameters are given in Table~\ref{tab:gpr-params}. In our work we $z$-normalize the prediction targets to have 0 mean and unit variance. This normalization is reversed when performing predictions.

\begin{table}[H]
\centering
    \begin{tabular}{ll}
    \toprule
        Hyperparameter & Setting \\\midrule
        $A$ & 4.52 \\
        $l$ & 4.54 \\
        $\sigma_n$ & 0.198 \\\bottomrule
    \end{tabular}
    \caption{Hyperparameter settings for kernels used in GPR model.}
    \label{tab:gpr-params}
\end{table}

\subsubsection{XGBoost}
We use the scikit-learn estimator interface provided in the DMLC implementation of XGBoost with the hyperparameter settings given in Table~\ref{tab:xgb-params}. The \texttt{random\_state} value is provided for reproducibility, since random initialization is used during training for sub-sampling feature columns.  
\begin{table}[H]
\centering
    \begin{tabular}{ll}
    \toprule
    Hyperparameter & Setting \\\midrule
        \texttt{n\_estimators} & \texttt{250} \\
        \texttt{max\_depth} & \texttt{7} \\
        \texttt{eta} & \texttt{0.05} \\
        \texttt{colsample\_bytree} & \texttt{0.5} \\
        \texttt{gamma} & \texttt{0.05} \\
        \texttt{random\_state} & \texttt{42} \\\bottomrule
    \end{tabular}
    \caption{Hyperparameter settings for XGBoost model.}
    \label{tab:xgb-params}
\end{table}

\subsubsection{SISSO}
\begin{figure}[H]
    \centering
    \includegraphics{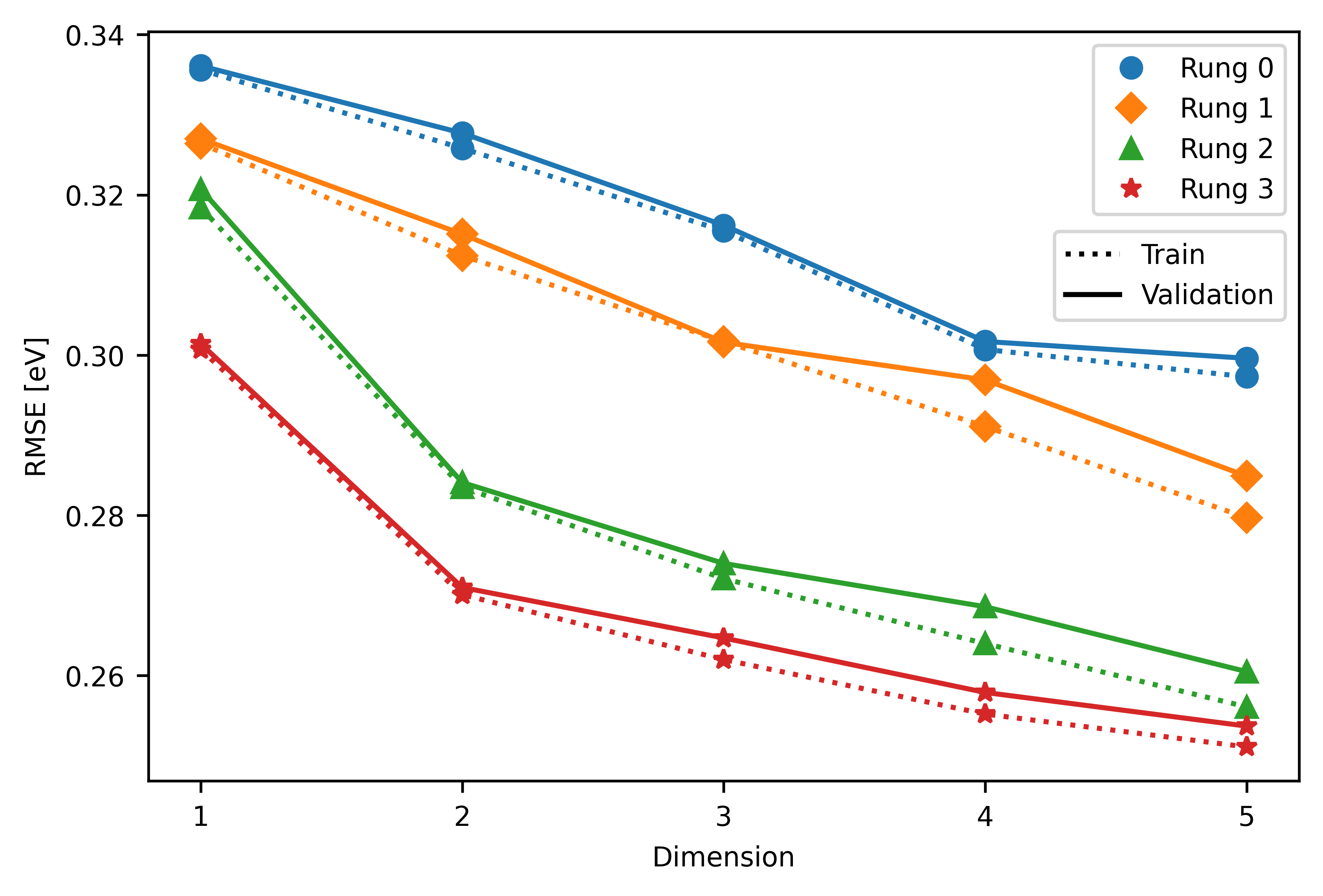}
    \caption{RMSEs from cross-validation to determine suitable hyperparameters for SISSO model.}
    \label{fig:sisso_hyperparams}
\end{figure}

We use the Python interface to SISSO provided by the pySISSO package. Figure~\ref{fig:sisso_hyperparams} shows the results of 5-fold cross validation on the training--validation set to determine a suitable rung and dimension for the SISSO model. 
To keep the computational cost tractable, we did not consider rungs higher than 3 and dimensions higher than 5. 
It is possible that the predictive performance would improve further at higher dimensions and rungs. 
For descriptor identification we set the sure independence screening (SIS) parameter to 10, and we make use of the following operator set: $\{+,-,\cdot,\div,x^{-1},x^{2},\sqrt x,\sqrt[3] x,\log x,|x| \}$.

Training of rung 3 SISSO models is done with a reduced set of primary features and operators. The features included in this reduced set are the 11 features with the highest importance score of the features according to the XGBoost model, as shown in Figure \ref{fig:fi_all}. The reduced operator set includes only operators that appear in the 5-dimensional rung 2 SISSO descriptor, and is as follows: $\{+,-,\cdot,\div,\log x,|x| \}$.

This reduction in primary features and operators is done in order to keep the run-time for the rung 3 feature space construction reasonably low.

\subsubsection{WWL-GPR}
The WWL-GPR model works with graph data of the atomic structures. In a graph, the nodes represent the atoms, and the edges the chemical bonds. Each node is decorated with a feature vector including the features denoted as “electronic” in Table S1, plus the SOAP descriptor. Atom-specific features (e.g., Pauli electronegativity) are not averaged over the binding site atoms as done when creating the features for the non-graph models, but the value calculated for each atom is attributed to the corresponding graph node. System-specific features (e.g., work function), denoted as “global” in Table S1, are attributed to all graph nodes corresponding to surface atoms. The information on the pDOS of C, H, and O atoms is excluded, as these atoms do not have a d-band. For missing feature values, a value of 0 is used.

Resembling message passing graph neural networks, the node features are propagated to the neighboring atoms with one step of the Weisfeiler-Lehman refinement scheme. The magnitude of this feature propagation is controlled by edge weights, here distinguished by the atoms that they connect: two surface atoms (\texttt{edge\_s\_s}), one surface atom and one adsorbate atom (\texttt{edge\_s\_a}), or two adsorbate atoms (\texttt{edge\_a\_a}). Then, the similarity between graphs is calculated by evaluating the Wasserstein distance (earth movers distance) between the graphs. Different contributions of each node during the calculation of the Wasserstein distance are controlled by the parameters \texttt{inner\_weight} and \texttt{outer\_weight}. We used fixed values of \texttt{outer\_cutoff} and \texttt{inner\_cutoff}, equal to 1 and 2, hence \texttt{inner\_weight} controls the weight of the nodes (atoms) involved in the bonds between the adsorbate and the surface atoms, while \texttt{outer\_weight} controls the weight of the nodes directly connected to those atoms.

Other hyperparameters of the WWL-GPR model are the GPR length scale (\texttt{gpr\_len}), which controls how quickly the correlation between function values decays with distance in the input space, and regularization (\texttt{gpr\_reg}), which controls the noise level of the GPR model.

We calculate the SOAP features of the atoms of the clean catalyst surface and the adsorbate in the gas phase with the Dscribe package \cite{himanen2020dscribe}. Following the methodology of Xu et al. \cite{Xu2022}, we do not distinguish between different chemical elements in the local atomic environments, as this information is already encoded in the other features. The parameters \texttt{r\_cut}, \texttt{n\_max}, \texttt{l\_max}, and \texttt{sigma} are set to 3, 4, 6, and 0.35, respectively.

\begin{table}[H]
\centering
    \begin{tabular}{ll}
    \toprule
    Hyperparameter & Setting \\
    \midrule
    \texttt{inner\_cutoff} & \texttt{1} \\
    \texttt{outer\_cutoff} & \texttt{2} \\
    \texttt{inner\_weight} & \texttt{0.74} \\
    \texttt{outer\_weight} & \texttt{0.14} \\
    \texttt{edge\_s\_s} & \texttt{0.87} \\
    \texttt{edge\_s\_a} & \texttt{0.65} \\
    \texttt{edge\_a\_a} & \texttt{0.25} \\
    \texttt{gpr\_reg} & \texttt{0.0092} \\
    \texttt{gpr\_len} & \texttt{30} \\
    \bottomrule
    \end{tabular}
    \caption{Hyperparameter settings for WWL-GPR model.}
    \label{tab:wwl-params}
\end{table}

We evaluated the WWL-GPR model without replacing geometric and stoichiometric features with the SOAP descriptor (hence, using only the features in Table S1). We observed a slight decrease in predictive performance compared to the version using SOAP, as shown in Table S5.

\begin{table}[H]
\centering
\begin{tabular}{lcc cc}
\toprule
 & \multicolumn{2}{c}{SOAP descriptor} & \multicolumn{2}{c}{Geometrical features} \\
\cmidrule(lr){2-3} \cmidrule(lr){4-5}
 & RMSE [eV] & MAE [eV] & RMSE [eV] & MAE [eV] \\
\midrule
WWL-GPR & 0.174 & 0.122 & 0.186 & 0.131 \\
\bottomrule
\end{tabular}
\caption{Root mean square error (RMSE) and mean absolute error (MAE) for the WWL-GPR model trained on the total training–validation dataset, using SOAP features, or stoichiometric and geometric features.}
\label{tab:wwl-params}
\end{table}

\subsection{Computational runtimes}

The four different ML models have different computational requirements for both training and evaluation.
To compare the four different models, we timed the training and evaluation procedures.
All timing was performed on a single core of an Intel Xeon Platinum 8358 CPU, as not all model implementations benefit from multi-core parallelization.
Training consisted of training a single model with optimized hyperparameters on the full training--validation split as outlined in Section II\,E\,5.
Evaluation consisted of using a trained model to predict adsorption energies of the testing split as outlined in Section II\,E\,5.
The runtimes are shown in Table~\ref{tab:runtimes}.

\begin{table}[H]
    \centering
    \begin{tabular}{lcc}
        \toprule
        & \multicolumn{2}{c}{Runtime} \\
        \cmidrule{2-3}
        Model & Training & Evaluation \\
        \midrule
        RBF-GPR & \qty{15.5}{\second} & \qty{98.8}{\milli\second} \\
        XGBoost & \qty{0.308}{\second} & \qty{1.68}{\milli\second} \\
        SISSO   & \qty{12.5}{\hour} & \qty{0.452}{\milli\second} \\
        WWL-GPR & \qty{59.3}{\minute} & \qty{19.6}{\minute} \\
        \bottomrule
    \end{tabular}
    \caption{Training and evaluation runtimes for the four models.}
    \label{tab:runtimes}
\end{table}

Note that further code optimization may reduce the runtimes, e.g., for the WWL-GPR model. For SISSO, we note that there exists a recent C++ implementation \cite{Purcell2022}, which may have a different runtime from the older Fortran implementation used here~\cite{sisso}.

For comparison, the DFT relaxation of a single initial binding configuration to a force convergence criterion of \qty{0.05}{\eV / \angstrom} and the subsequent refining calculation with a (\numproduct{2x2}) \textit{k}-point grid required on average 389 core hours.

\clearpage

\section{Comparison to foundational models}

Given the recent rise of foundational pre-trained models (FPMs), we provide in this section the methodology and results of three tests performed to evaluate the effectiveness of using FPMs to predict adsorption energies. The FPMs used in our tests are CHGNet and MACE-MP-0a, where we use the medium model size for MACE-MP-0a.
For the tests, we randomly sample a subset of binding configurations for which we have performed structure relaxation with DFT.

In the first two tests, we then further relax these structures---as well as the corresponding clean surfaces and the gas-phase formate molecule---in the FPM potential using the same force convergence criterion as used in the DFT relaxation. Here we term this additional relaxation with the FPM as a `re-relaxation.' We chose to use the DFT-relaxed structure, and not the initial binding configuration obtained as described in Section II\,B of the main article, as the starting geometry for the re-relaxation to increase the likelihood that the formate molecule remains at the same adsorption site so that a meaningful comparison of the DFT and FPM adsorption energy can be performed. In about 10 (8) out of 60 cases of re-relaxation in the MACE-MP-0a (CHGNet) potential do we see that the adsorbate nevertheless moves to a different site than the site of the DFT-optimized structure.

\begin{figure}[H]
    \centering
    \includegraphics{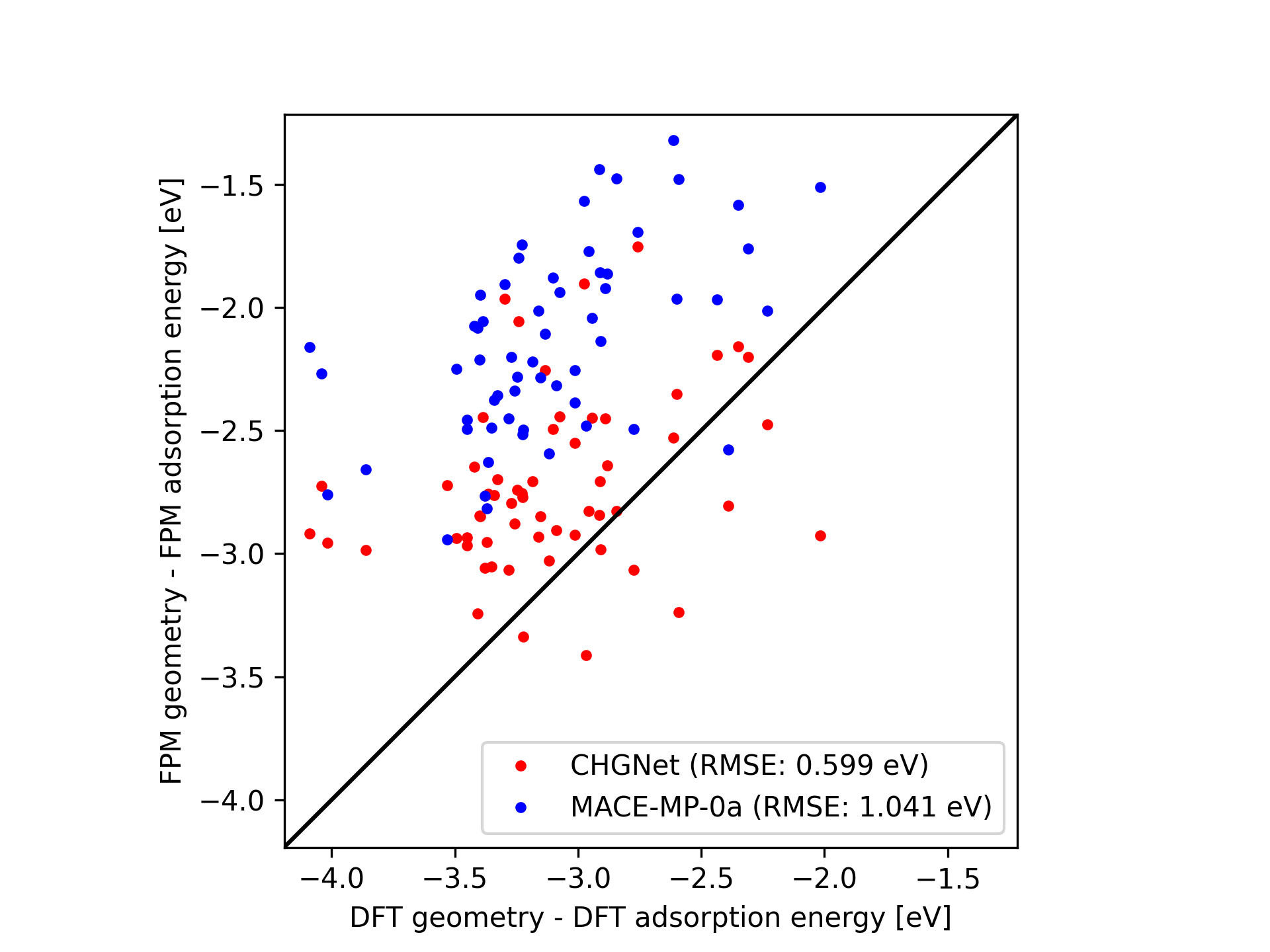}
    \caption{Parity plot of adsorption energies obtained with the FPM potential, where all geometries have been relaxed in this potential, versus corresponding results for DFT.}
    \label{fig:parity-dftgeo-mlgeo+mlsp}
\end{figure}

\begin{figure}[H]
    \centering
    \includegraphics{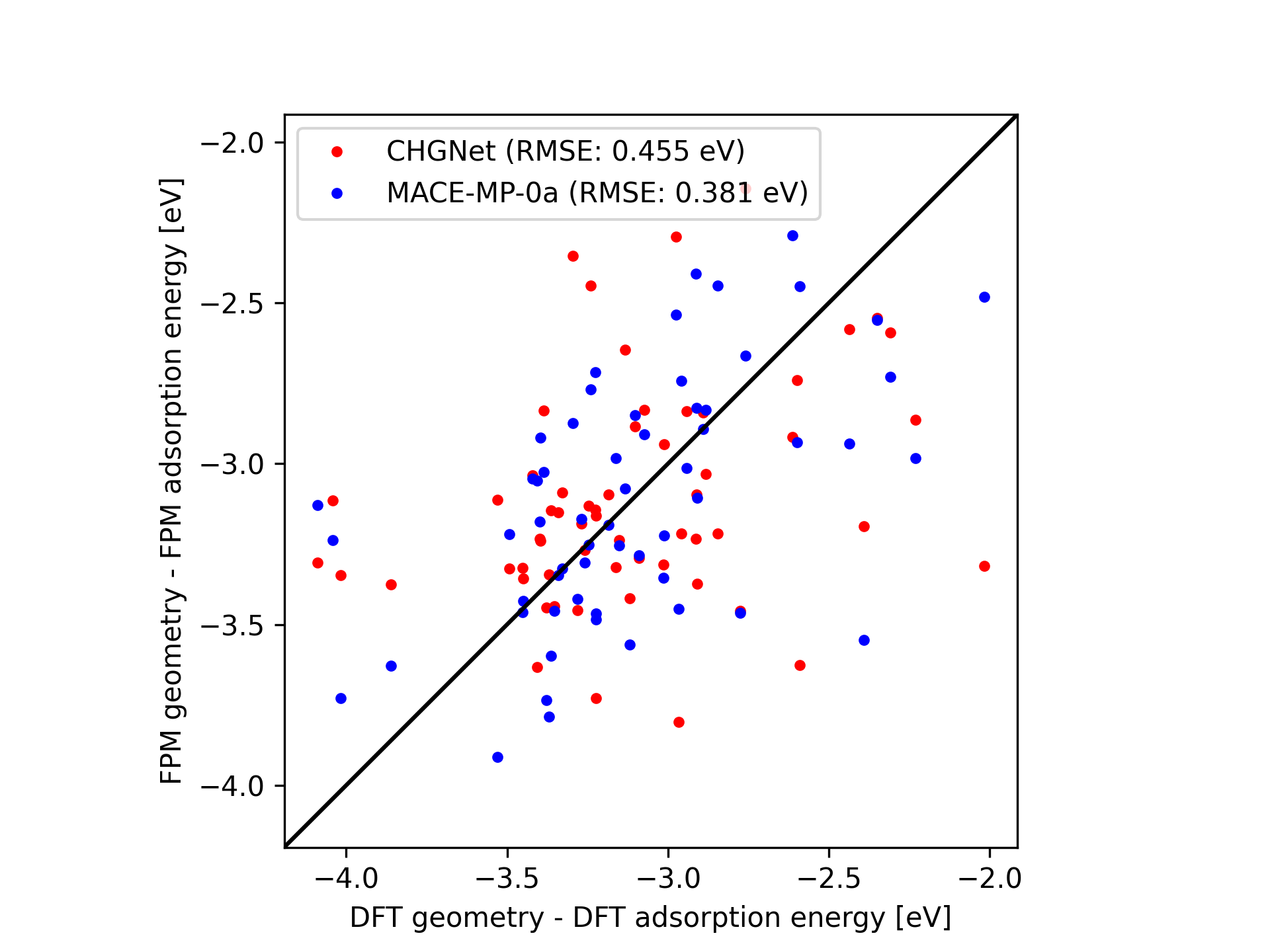}
    \caption{Same as Figure \ref{fig:parity-dftgeo-mlgeo+mlsp}, but where the formate energy term in the FPM-evaluated adsorption energies has been left as a free fitting parameter.}
    \label{fig:parity-dftgeo-mlgeo+mlsp-formatefit}
\end{figure}

In the first test, we compute the FPM adsorption energy with all geometries being relaxed in the FPM potential. The parity plot comparing this approach against DFT is shown in Figure~\ref{fig:parity-dftgeo-mlgeo+mlsp}. The very high RMSEs obtained and the systematic deviation with respect to DFT may in part be due to the FPM models not describing the formate molecule well, since they have been primarily trained on inorganic materials. To account for this, we provide in Figure~\ref{fig:parity-dftgeo-mlgeo+mlsp-formatefit} the corresponding parity plot, where the energy of gas-phase formate in the FPM binding energy expression has been left as a free fitting parameter. However, the RMSEs remain very large, around \qty{0.4}{\eV} for both FPMs, which is substantially larger than the RMSE around \qty{0.2}{\eV} obtained with the best ML models in the main paper (cf.\ Table~III).

\clearpage

\begin{figure}[H]
    \centering
    \includegraphics{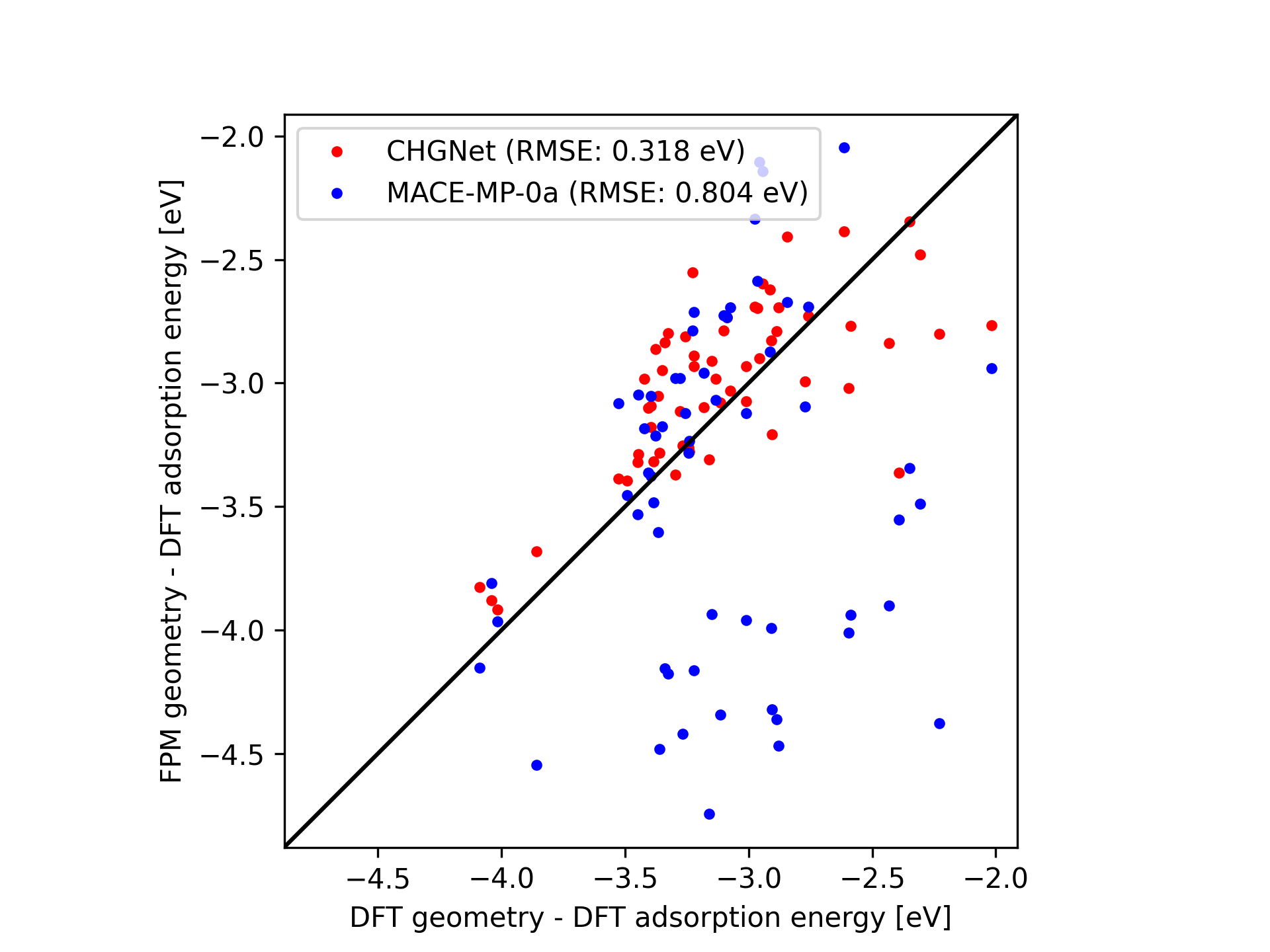}
    \caption{Parity plot of DFT-calculated adsorption energies for structures relaxed in the FPM potential versus corresponding energies for structures relaxed with DFT.}
    \label{fig:parity-dftgeo-mlgeo}
\end{figure}

In the second test, we investigate the quality of the structure of the binding configuration obtained through re-relaxation in the FPM potential, as described above. That is, we carry out a DFT single-point evaluation of the energy of this structure as well as of the FPM-optimized structures of the clean surface and gas-phase formate. The parity plot comparing these adsorption energies against those calculated using DFT-optimized structures is shown in Figure \ref{fig:parity-dftgeo-mlgeo}. In this case, CHGNet seems to provide significantly better structures than MACE-MP-0a. However, even for CHGNet, the RMSE obtained remains substantially higher than the model results presented in Table~III in the main paper.

\clearpage

\begin{figure}[H]
    \centering
    \includegraphics{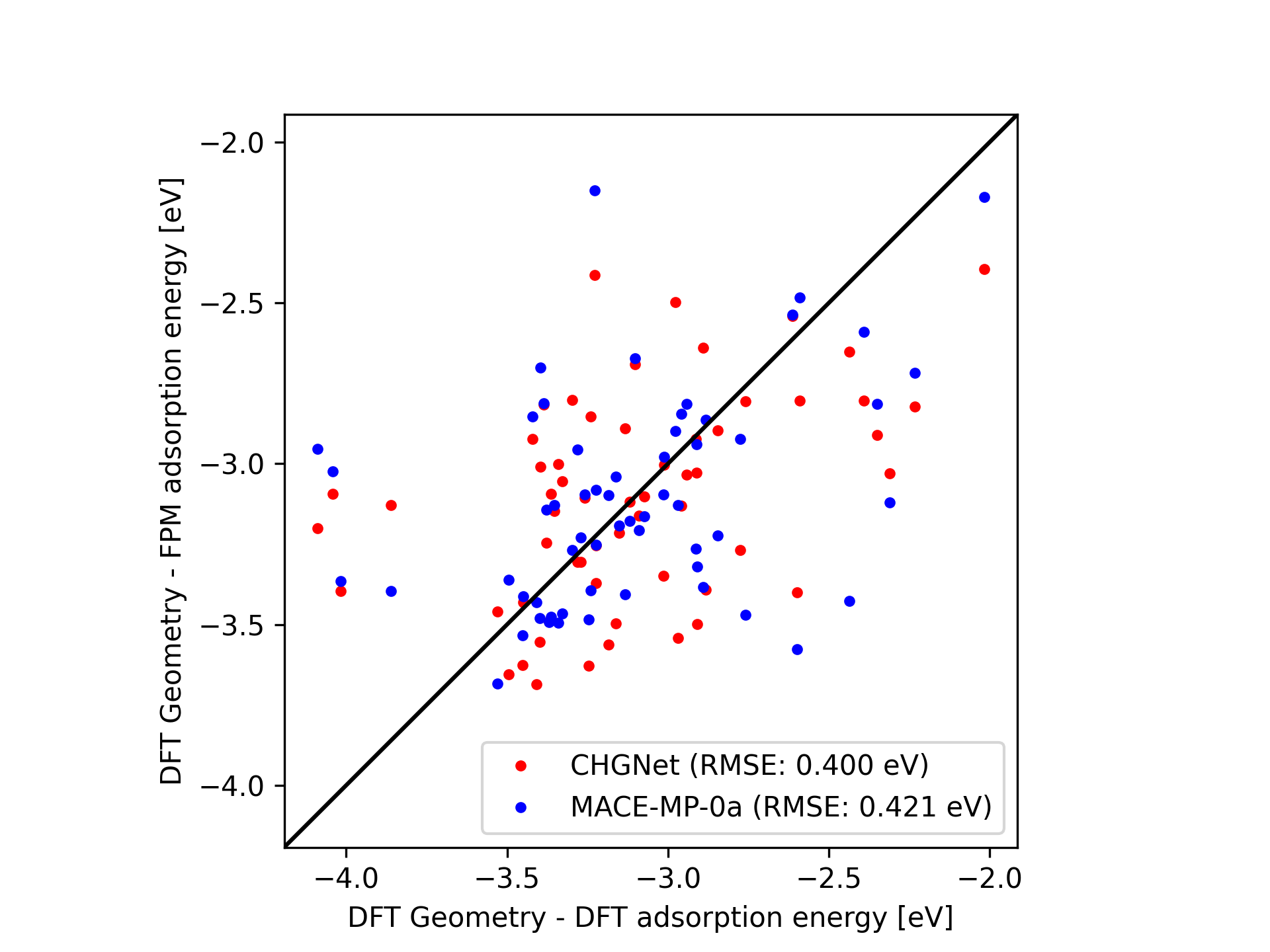}
    \caption{Parity plot of FPM-calculated adsorption energies using DFT-relaxed structures versus DFT-calculated adsorption energies with the same structures.}
    \label{fig:parity-dftgeo-mleval}
\end{figure}

Finally, in the third test, we investigated adsorption energies calculated with the FPM potential, but using structures optimized with DFT. That is, we carry out FPM single-point energy evaluations of the DFT-optimized structures of formate on the cluster and the clean cluster. Again, we leave the FPM energy of the gas-phase formate molecule as a free fitting parameter. The FPM adsorption energies obtained with this approach are compared to the DFT adsorption energies (calculated at the same geometry) in Figure~\ref{fig:parity-dftgeo-mleval}. Also in this case do we see very large RMSEs (around \qty{0.4}{\eV}) with the FPM models.

Overall, we conclude from these tests that FPM models are not competitive with the models presented in the main paper for prediction of adsorption energies. Fine-tuning may improve their performances, however, this is out of scope of the present work.

\printbibliography